\documentclass[nofootinbib,floats,superscriptaddress,eqsecnum,tightenlines,11pt]{revtex4}

\usepackage{hyperref}
\usepackage{graphicx}
\usepackage{amsmath,amssymb,amsfonts,amsthm,latexsym,stmaryrd}
\usepackage{marginnote}
\usepackage{color}

\def\be{\begin{equation}}
\def\ee{\end{equation}}
\def\ba{\begin{eqnarray}}
\def\ea{\end{eqnarray}}
\def\bas{\begin{subequations}\begin{eqnarray}}
\def\eas{\end{eqnarray}\end{subequations}}

\def\tr{\text{tr}}

\def\su{\mathfrak{su}}
\def\so{\mathfrak{so}}

\begin{document}

\title{Translation invariant time-dependent massive gravity:\\ 
Hamiltonian analysis}

\author{Jihad Mourad}
\email{mourad@apc.univ-paris7.fr}
\affiliation{Laboratoire APC -- Astroparticule et Cosmologie, Universit\'e Paris Diderot, 75013 Paris, France}
\author{Karim Noui}
\email{karim.noui@lmpt.univ-tours.fr}
\affiliation{Laboratoire de Math\'ematiques et Physique Th\'eorique, Universit\'e Fran\c cois Rabelais, Parc de Grandmont, 37200 Tours, France}
\affiliation{Laboratoire APC -- Astroparticule et Cosmologie, Universit\'e Paris Diderot, 75013 Paris, France}
\author{Dani\`ele A.~Steer}
\email{steer@apc.univ-paris7.fr}
\affiliation{Laboratoire APC -- Astroparticule et Cosmologie, Universit\'e Paris Diderot, 75013 Paris, France} 

\date{\today}

\begin{abstract}
The canonical structure of the massive gravity in the first order moving frame formalism
is studied. We work in the simplified context of translation invariant fields,
with mass terms given by general non-derivative interactions, invariant under the diagonal Lorentz group, depending  on the moving frame as well as a fixed reference frame.   We prove that the only mass terms which give 5 propagating degrees of freedom are the dRGT mass terms, namely those which are linear in the lapse. We also complete the Hamiltonian analysis with  the dynamical evolution of the system.
\end{abstract}

\maketitle

\tableofcontents
\section{Introduction}

General relativity is the interacting theory of a massless spin two particle --- the graviton --- and the consistency of these interactions is guaranteed by  diffeomorphism invariance, which also provides general relativity with its geometrical interpretation.  Local reparametrisation invariance leads to the Einstein equations, and also to four first class constraints.  Thus the covariant symmetrical tensor with ten components (in 4 dimensions) describes the 2 degrees of freedom of the graviton.  

As its long history shows, the study of small deviations of general relativity is of great interest, both from a theoretical and phenomenological point of view, see  e.g.~\cite{Rubakov:2008nh,Hinterbichler:2011tt,claudia} and references within. 
A generic massive deformation of the linearized Einstein equations, however, gives rise to 6 propagating degrees of freedom corresponding to those of a massive spin two field and a scalar (the four first class constraints which are inherited from the Einstein equations become second class: see below for a detailed discussion of the constraints algebra). In addition, the scalar couples to matter with a propagator having a sign characteristic of a pathological ghost, the Boulware-Deser ghost \cite{Boulware:1973my}. 

In the linearised theory, Fierz and Pauli \cite{Fierz:1939ix} showed that when 
the mass term is a particular combination of the two quadratic invariants, a new constraint arises which removes the extra degree of freedom. However, this combination is not a consequence of an additional gauge symmetry, and hence a generic self-interaction reintroduces the extra degree of freedom, as was confirmed by Boulware and Deser \cite{Boulware:1973my}. From a Hamiltonian point of view, the Fierz-Pauli combination is the unique one linear in the lapse (or the time-time component of the rank two tensor), and we shall come back to this important point later in the paper.
From a phenomenological point of view, the massive deformation  which seemed excluded due to the van dam Zakharov discontinuity  characteristic of the linear Fierz-Pauli theory \cite{vanDam:1970vg}
was revived by Vainstein \cite{Vainshtein:1972sx} who argued that non linearities cannot be neglected even at the solar system distance scales (for a recent  review see \cite{Babichev:2013usa}).

In a seminal paper \cite{deRham:2010kj}, de Rham, Tolley and Gabadadze proposed a ghost-free non-linear extension of the Fierz-Pauli action. The absence of the ghost was first shown in the decoupling limit \cite{Creminelli:2005qk,deRham:2010ik,ArkaniHamed:2002sp}, whilst the full Hamiltonian analysis was discussed in \cite{Hassan:2011ea,Hassan:2011zd,Hassan:2011tf,Hassan:2011hr,Kluson,OthersCounting,Alexandrov:2013rxa}.\footnote{and debated in \cite{Alberte:2010qb,Chamseddine:2011mu,Chamseddine:2013lid}.}
In its original version, the theory was formulated with two metrics \cite{Isham:1971gm} one of which, say $h$ can be fixed and the the other one, $g$ describes the massive graviton, and the mass terms are built from the  matrix square root of $g^{-1}h$.  In general this is not necessarily well defined (nor indeed real) \cite{dmz1,dmz2}, but its existence can be imposed if a certain symmetry condition \cite{Chamseddine:2011mu,Volkov:2012wp,dmz2} is satisfied. (We will discuss the origin of this condition in the Hamiltonian framework below.) In that case, the theory takes a rather simpler form in a moving basis or veilbein formulation \cite{Hinterbichler:2012cn,Nibbelink:2006sz} where the terms become polynomial and where, for certain mass terms, the additional constraint removing the Boulware-Deser ghost is easily obtained \cite{dmz1}.  It is precisely this vielbein formulation which we consider in this article. 

Here we perform the Hamiltonian analysis of  general translation-invariant fields \cite{ms1,ms2} to further explore the vielbein theory in the first order formalism; that is when the moving frame and the spin connexion are considered as independent dynamical variables. First, we consider the most general nonlinear mass term with global Lorentz symmetry and show that when this mass term is not linear in the lapse function then additional degrees of freedom are present.  Thus it follows that also in the general case of space- and time-dependent fields, there will be additional degrees of freedom when the mass term is not linear in the lapse.   The constraints analysis also enables us to determine the origin of the symmetry condition mentioned above (see also \cite{Alexandrov:2013rxa}). Returning to translation invariant fields, we then consider the most general mass terms linear in the lapse function (which coincide with the dRGT mass terms \cite{deRham:2010kj}), determine all the constraints, show that the degrees of freedom are those of a massive spin two field and finally write the time evolution equations. The framework we consider of time-dependent fields is a simplified one, but it has the advantage of allowing one to find all the constraints and to compute explicitly the Dirac matrix together with the Dirac bracket (which is equivalent to solving the constraints), contrary to the case of the Hamiltonian treatment of a general space and time dependent field where the constraints cannot be explicitly solved \cite{Hassan:2011ea,Hassan:2011zd,Hassan:2011tf,Hassan:2011hr,Kluson,OthersCounting,Alexandrov:2013rxa}.  Notice, however, that in \cite{Alexandrov:2013rxa} it was shown that the dRGT theory is ghost free in the general case of two space- and time-dependent vierbeins. Finally, we also note that the work presented here extends that of \cite{ms1,ms2} which considered time-dependent fields, but only for two specific mass terms: here we give the equations of motion for the general mass term.

The plan of this paper is as follows.
We begin, in section \ref{section2}, by recalling the main steps and the main results of the  Hamiltonian analysis of general relativity in the first order formalism.
This will allow us to introduce notation which will be used throughout the paper.
In section \ref{section3} we construct the most general mass terms for massive gravity, discuss their symmetries, and decompose them in their ADM form, thus setting the scene for section \ref{section4} where we carry out an ADM analysis of these most general mass terms.  We conclude that unless the mass terms take the specific dRGT form \cite{deRham:2010kj}, the Boulaware-Deser ghost is present. In section \ref{section5} we carry out a complete Hamiltonian analysis of the dRGT massive gravity for translation invariant fields.  In section \ref{section6} we determine the equations of motion through the calculation of the Dirac bracket, and thus obtain the time evolution of all the dynamical variables. Finally, we collect our conclusions in section \ref{section7} and the technical details in the Appendices.

\section{Hamiltonian analysis of pure gravity}
\label{section2}

We begin by recalling (the main steps of) the Hamiltonian analysis of the pure gravity action in four space-time dimensions, in the first order formalism. Here
 the gravitational dynamical variables are a tetrad field $\theta$ and a spin-connection $\omega$, which are a priori independent one-forms
 taking values respectively in  flat Minkowski space-time and in the Lorentz algebra $\so(1,3)$. 
Their components are denoted by 
$$\theta^I=\theta_\mu^I dx^\mu \qquad {\rm and}  \qquad \omega^{IJ}=\omega_\mu^{IJ} dx^\mu$$ 
where greek letters $\mu,\nu,\cdots \in \{0,1,2,3\}$ refer to space-time indices, and capital latin letters $I,J,\cdots \in \{0,1,2,3\}$ to internal Lorentz (or flat) indices which are raised and lowered by the flat Minkowski metric $\eta=\text{diag}(-1,+1,+1,+1)$. 

The pure gravity action is given by
\begin{eqnarray}\label{puregravity}
S[\theta,\omega]  =   \frac{1}{8}   \int \epsilon_{IJKL}   \theta^I \wedge \theta^J \wedge F^{KL} \label{gravity action}
\end{eqnarray}
where $\epsilon$ is the totally antisymmetric tensor with the convention $\epsilon_{0123}=1$, and the $\so(1,3)$-valued 2-form
$F$ is the curvature 2-form of the spin-connection $\omega$ with components
\begin{eqnarray}
F_{\mu \nu}= \partial_\mu \omega_\mu - \partial_\nu \omega_\mu + [\omega_\mu,\omega_\nu].
\label{Fdef}
\end{eqnarray}
For the Hamiltonian analysis, as usual we consider a space-time $\cal M$ of the form $\Sigma \times T$ where $\Sigma$
is three dimensional space and $T \subset \mathbb R$ indicates the time direction. 

Before proceeding it is useful to introduce the following notation which simplify matters considerably. 
We use lower case latin letters from the beginning of the alphabet $a,b,\cdots \in \{1,2,3\}$ to denote space indices, while lower case latin letters from the
middle of the alphabet  $i,j,k,\ell \cdots \{1,2,3\}$ denote space-like Lorentz indices. Furthermore we distinguish between the boost and rotational components of the spin-connection as follows
\begin{eqnarray}
A_\mu^i = \omega_\mu^{0i} \;\;\;\;\; \text{and} \;\;\;\; \omega_\mu^i = \frac{1}{2} \epsilon^{ijk} \omega_\mu^{jk} \;\;\; \text{with} \; \epsilon^{ijk}=\epsilon^{0ijk}.
\end{eqnarray}
These two components will play very different r\^oles in the Hamiltonian dynamics of gravity. 
It will be useful to consider the $A_\mu^i$ and $\omega_\mu^i$ as components of the 3-dimensional vectors, $\vec{A}_\mu$ and $\vec{\omega}_\mu$ on the flat Euclidean space $\mathbb R^3$ (below we will omit the arrows).
In terms of these vectors,  a straightforward calculation shows that the components
of the curvature 2-form (\ref{Fdef}) are given by
\begin{eqnarray}
F_{\mu\nu}^{0i} & = & \partial_\mu A_\nu^{i} - \partial_\nu A_\mu^i + (\omega_\nu \times A_\mu)^i - (\omega_\mu \times A_\nu)^i \\
F_{\mu\nu}^i & \equiv & \frac{1}{2} \epsilon^{i}{}_{jk} F_{\mu\nu}^{jk}=\partial_\mu \omega_\nu^{i} - \partial_\nu \omega_\mu^i + (A_\mu \times A_\nu)^i - (\omega_\mu \times \omega_\nu)^i 
\end{eqnarray}
where  $\times$ denotes the vector product between 3-dimensional vectors defined by $(u\times v)^i=\epsilon^{i}{}_{jk}  u^j v^k$ for any vectors $u,v \in \mathbb R^3$. 
In the following we also use the scalar product notation $u \cdot v=u_iv_i$.
For simplicity, we will denote $F_{\mu\nu}$ the vector with components $F_{\mu\nu}^{0i}$, and $G_{\mu\nu}$ the vector whose components are $F_{\mu\nu}^i$.

It is also convenient to separate the tetrad fields $\theta^I_\mu$ into their pure space-like components 
$\theta_a^i$, as well as the time-like $\theta_0^0$ and mixed components $\theta_0^i$ and $\theta_a^0$. We will use the following notation:
\begin{eqnarray}\label{tetrad components}
E^a_i \equiv \frac{1}{2} \epsilon^{abc}\epsilon_{ijk} \theta_b^j \theta_c^k \;\;\;,\;\;\; \theta_0^0 \equiv N \;\;\;,\;\;\; \theta_0^i \equiv N^a \theta_a^i \;\;\;,\;\;\; \theta_a^0\equiv \theta_a^i \chi_i \;,
\end{eqnarray}
thus defining two new vectors in $\mathbb R^3$, namely $\chi$ and $\theta_0$ and a 3 dimensional matrix $E$. From now on, we will use the notation $\theta$ for the three-dimensional space matrix 
(that is $\theta^i_a$ viewed as a $3\times 3$ matrix) and
$\theta_{(4)}$ the four-dimensional one.  The variables $N$ and $N^a$ are the well-known  lapse function and the shift vector of gravity. 
Finally, when $\theta$ is invertible,
then $E$ is related to the inverse of $\theta$ by
\begin{eqnarray}\label{inverse E}
E \equiv \vert \theta \vert \, \theta^{-1} \;\;\; \text{where} \;\;\; \vert \theta \vert \equiv \text{det}(\theta)=\vert E \vert^{1/2}.
\end{eqnarray}
From now on we assume that  $\theta$, hence $E$, is indeed invertible (a necessary requirement for first order gravity to be equivalent to the standard second order formulation of gravity).
We choose $\vert E \vert >0$ without loss of generality. 

The Lagrangian density in (\ref{puregravity}) can then be rewritten in the form
\begin{eqnarray}
{\cal L}_{grav} & = & E_a \cdot (\partial_0 A_a - \partial_a A_0 + \omega_a \times A_0 - \omega_0 \times A_a) \nonumber \\
 &  + &  \chi \times E^a \cdot (\partial_a \omega_0 - \partial_0 \omega_a + A_a \times A_0 - \omega_a \times \omega_0)\nonumber \\
 &  + & \frac{1}{2} N \frac{E^a \times E^b}{\vert E \vert^{1/2}} \cdot G_{ab} + N^b E^a \cdot F_{ab} -  \frac{1}{2} \epsilon^{abc} N^d \theta_a^l \theta_{di} \chi_l G_{bc}^i.
 \label{step1}
\end{eqnarray}
This can be simplified further by introducing a new lapse  function $\cal N$ and a new shift vector 
${\cal N}^a$, following \cite{Alexandrov:2000jw}, which are linear combinations of the original ones  (\ref{tetrad components}):
\begin{eqnarray}\label{new lapse}
N= {\cal N} + (\theta_a \cdot \chi) {\cal N}^a \;\;\;\text{and}\;\;\;
N^a={\cal N}^a + \frac{E^a\cdot \chi}{\vert E \vert^{1/2}} {\cal N}.
\end{eqnarray}
In the time gauge ($\chi=0$), they reduce to the usual lapse and shift. In absence of time gauge, and up to an irrelevant total derivative term, (\ref{step1}) becomes
\begin{eqnarray}\label{Lag density}
{\cal L}_{grav}  =  E_a \cdot  \partial_0 A_a - (\chi \times E^a) \cdot \partial_0 \omega_a + A_0\cdot U + \omega_0 \cdot S +  \frac{\cal N}{2\vert E \vert^{1/2}}  H^{grav}+  {\cal N}^a H^{grav}_a 
\end{eqnarray}
where 
\begin{eqnarray}
 U & = & \partial_a E^a + E^a \times \omega_a +(\chi \times E^a)\times A_a  ,\label{U}\\
 S & = & \partial_a(E^a\times \chi) - (\chi \times E^a)\times \omega_a +E^a \times A_a , \label{S}\\
 H^{grav}& = &  (E^b\cdot \chi)  ( E^a\cdot F_{ab}) +[E^a\times E^b - (E^a\times E^b \cdot \chi) \chi] \cdot G_{ab} ,\label{H} \\
 H^{grav}_a& = &E^b \cdot F_{ba} + (E^b \times \chi) \cdot G_{ba}.\label{Ha}
\end{eqnarray}

From (\ref{Lag density}), the Hamiltonian structure of pure gravity becomes clear: the dynamical variables are a priori given by the components $A_a$
and $\omega_a$ of the spin connection together with their conjugate momenta which satisfy the Poisson brackets
\begin{eqnarray}\label{grav poisson}
\{ E^a_i(x),A_b^j(y)\} = \delta^a_b \delta_i^j \, \delta(x-y) \;\;\;\; \text{and} \;\;\;\; \{P^a_i(x),\omega_b^j(y)\}= \delta^a_b \delta_i^j \, \delta(x-y)\;.
\end{eqnarray}
Here
\begin{eqnarray}\label{constraintonP}
P^a_i \equiv \frac{\delta {\cal L}_{grav}}{\delta \omega_a^i} = - ( \chi \times E^a)_i,
\end{eqnarray}
and this is clearly not totally independent of $E$ -- the $P^a_i$ therefore satisfy some constraints which we discuss below.
 The remaining variables, $A_0$, $\omega_0$, $\cal N$ and ${\cal N}^a$ are Lagrange multipliers  which enforce $3+3+1+3=10$
constraints. It can be shown that (up to adding second class constraints to them) they form a set of first class constraints and therefore generate the local symmetries
of the theory:   the local Lorentz invariance (6 dimensional symmetry) and the space-time diffeomorphism invariance (4 dimensional symmetry) \cite{Peldan:1993hi}. 

The system admits also second class constraints. The first set come directly from the expression of the Lagrangian density (\ref{Lag density}) (and therefore are primary second
class constraints), and enforce that the $P$ variables are not independent of $E$, (\ref{constraintonP}). More precisely, only three components out of the nine are  independent, and as a consequence, the $P$ variables satisfy a set of 6 constraints which can be formulated as follows:
\begin{eqnarray}\label{Phiconstraint}
\Phi^{(ab)} = E^a \cdot P^b + E^b \cdot P^a =0 \,.
\end{eqnarray}
We used the notation $(ab)$ to make explicit that $\Phi^{(ab)}$ is symmetric.
It can be shown that such a set constraint is equivalent to saying that there exists a vector $\chi$ such that $P^a=E^a\times \chi$. These constraints are commonly called simplicity constraints.  
There are no more primary constraints. 

Time evolution is generated by the total Hamiltonian of the system $H_{tot}$, which as usual in any theory invariant under diffeomorphisms is a linear
combination of the primary constraints;
\begin{eqnarray}
H_{tot} & = & E_a \cdot  \partial_0 A_a + P^a \cdot \partial_0 \omega_a - {\cal L}_{grav}   \\
& = & -\left( A_0\cdot U + \omega_0 \cdot S +  \frac{\cal N}{2\vert E \vert^{1/2}}  H^{grav}+  {\cal N}^a H^{grav}_a + \mu_{ab} \Phi^{(ab)}\right)
\end{eqnarray}
Here the Lagrange multipliers $\mu_{ab}$ implement the constraint $\Phi^{(ab)}=0$. 
The stability under time evolution of the  first class constraints does not create secondary constraints, while the stability of the second class
constraints $\Phi^{(ab)}$ which can be shown to introduce a set of 6 new secondary constraints \cite{Peldan:1993hi}
\begin{eqnarray}\label{Psiconstraint}
\Psi^{(ab)}=\{H_{tot},\Phi^{(ab)}\}. 
\end{eqnarray}
It is not necessary to express explicitly these secondary constraints. 
The Dirac algorithm stops here, hence there is no tertiary constraints. An easy analysis shows that $\Psi^{(ab)}$ are also second class and therefore we end up with $10$ first class constraints
and $12$ second class constraints. As we started with $36$ non-physical degrees of freedom (\ref{grav poisson}), we finish with $36-2\times 10-12=4$ physical degrees of freedom in the phase
space as expected for gravity in four space-time dimensions.

\section{Massive gravity}
\label{section3}

In this section we introduce a mass term to the action (\ref{puregravity}).  First we construct the most general mass term out of two metrics, and discuss its symmetries.  We then focus on the case in which the second metric is fixed, and taken to be Minkowski, and finally we end the section with the ADM decomposition of this general mass term. Its Hamiltonian structure will then be studied in section \ref{section4} where we will see that the theory
 contains new physical degrees of freedom.

\subsection{Building blocks of the action}

We will focus on massive gravity built out of two metrics  $g_{\mu \nu}$ and $h_{\mu \nu}$ \cite{Isham:1971gm,Damour:2002ws}, one of which, say $h_{\mu \nu}$ can be non-dynamical.  The mass term in general bimetric theories depends on an invariant function ${\cal V}(g^{\mu \nu} h_{\mu \nu})$, so that the action is invariant under the diagonal diffeomorphism.  If, instead of the metrics, we use as dynamical variables the moving frames $\theta^A$ and $f^B$ defined by
\ba
&\eta_{AB}\theta^{A}{}_{\mu}\theta^{B}{}_{\nu}=g_{\mu\nu}\ ,
\nonumber
\\
&\eta_{AB}f^{A}{}_{\mu}f^{B}{}_{\nu}=h_{\mu\nu}\ ,
\nonumber
\ea
with their corresponding inverses
\be
\theta^{A}(e_{B})=\theta^{A}{}_{\mu}e_{B}{}^{\mu}=\delta^{A}{}_{B}, \qquad f^{A}(\ell_{B})=f^{A}{}_{\mu}\ell_{B}{}^{\mu}=\delta^{A}{}_{B}
\nonumber
\ee
then the most general mass term will be an invariant function of
\be
\theta^A(\ell_B) \equiv \Theta^{A}{}_{B}.
\nonumber
\ee
(Notice that this is independent of the spin-connexion.)
The resulting action is then invariant under the diagonal diffeomorphisms together with the diagonal local Lorentz group: indeed, under the latter transformation $\Theta^{A}{}_{B}$ transforms as 
$\Theta \mapsto \Lambda \Theta \Lambda^{-1}$.
The most general mass terms will then be constructed out of the invariants
\begin{eqnarray}
\label{hopeless}
\tilde{\varphi}_0 \equiv   \vert \Theta \vert,\qquad
\tilde{\varphi}_1 \equiv   \tr(\Theta), \qquad
\tilde{\varphi}_2 \equiv  \tr(\Theta^2),  \qquad
\tilde{\varphi}_3 \equiv  \tr(\Theta^3) , \qquad
\end{eqnarray}
and we denote the corresponding Lagrangian by ${\cal L}_m(\tilde{\varphi}_0,\ldots,\tilde{\varphi}_3)$.

In the remainder of this paper we fix $h_{\mu \nu} = \eta_{\mu \nu}$ and 
 \be
 f^I_\mu=\delta_\mu^I
 \label{fdelta}
 \ee
 thus working in cartesian coordinates.  Furthermore this choice enables us to identify the Lorentz internal indices ($I,J,\cdots$) with the space-time indices  ($\mu,\nu,\cdots$), so that in particular $\Theta$ is nothing other than $\theta_{(4)}$ defined in (\ref{tetrad components}).   The resulting action is then invariant under transformations
 \be
 x'=\Lambda^{-1}x \qquad {\rm and} \qquad \theta_{(4)} \longmapsto \Lambda \theta_{(4)} \Lambda^{-1} 
\ee
where $\Lambda$ is is now a global (space-time independent) Lorenz-transformation.   
Finally, we impose that the Minkowski moving frame $\theta^A = f^A = dx^A$ is a solution, and thus 
\be \theta_{(4)}=  \mathbb{I}.
\label{simp}
\ee
From the Einstein equations, this constrains the interaction terms to satisfy 
\be
\frac{\partial {\cal{L}}_m}{\partial \theta^{a}_{\mu}} (\theta_{(4)}=  \mathbb{I}) =0
\nonumber
\ee 
which, on using (\ref{hopeless}), reduces to
\be
\frac{\partial {\cal{L}}_m}{\partial \theta^{a}_{\mu}} (\theta_{(4)}=  \mathbb{I}) = \frac{\partial {\cal{L}}_m}{\partial \tilde{\varphi}_n} \frac{\partial {\cal{L}}_m}{\partial \theta^{a}_{\mu}} (\theta_{(4)}=  \mathbb{I})  = \mathbb{I} \left[  \frac{\partial {\cal{L}}_m}{\partial \tilde{\varphi}_0} + 2  \frac{\partial {\cal{L}}_m}{\partial \tilde{\varphi}_1} +  3 \frac{\partial {\cal{L}}_m}{\partial \tilde{\varphi}_2}+  \frac{\partial {\cal{L}}_m}{\partial \tilde{\varphi}_4}\right] (\theta_{(4)}=  \mathbb{I}) =0
\label{nearly}
\ee
where the condition (\ref{simp}) implies $(\tilde{\varphi}_0,\tilde{\varphi}_1,\tilde{\varphi}_2,\tilde{\varphi}_3)=(1,4,4,4)$.

\subsection{The most general mass term in its ADM form}

 In their 3+1 form, the basis functions of the algebra of Lorentz invariant functions (\ref{hopeless}) become (on using the decomposition in (\ref{tetrad components})),
\begin{eqnarray}
\tilde{\varphi}_0&=  &  \vert \theta \vert (N-N^a\theta_a^i\chi_i),\nonumber \\
\tilde{\varphi}_1& = &  N + \tr(\theta), \nonumber  \\ 
\tilde{\varphi}_2& = & N^2 + 2 \theta_0^a \theta^0_a + \tr(\theta^2),  \nonumber  \\
\tilde{\varphi}_3& = &  N^3 + 3N \theta_0^a \theta_a^0 + 3 \theta^0_a \theta^a_b \theta^b_0 + \tr(\theta^3) . \nonumber  
\end{eqnarray}
To simplify the canonical analysis,  it is more convenient to work with functions that are linear in the lapse $N$, and thus we replace $\tilde{\varphi}_{2,3}$ by 
\begin{eqnarray}
\varphi_2 &\equiv & -\frac{1}{2}(\tilde{\varphi}_2 - \tilde{\varphi}_1^2) \nonumber 
\\
 &= & N \tr(\theta) - \theta^0_a \theta^a_0 + \frac{1}{2} \left[ (\tr(\theta))^2 - \tr(\theta^2)\right] 
 \nonumber 
 \\
\varphi_3 & \equiv & \frac{2}{3}\tilde{\varphi}_3 -  \tilde{\varphi}_1 \tilde{\varphi}_2 + \frac{1}{3}\tilde{\varphi}_1^2
\nonumber 
  \\
&= &
N \left[ (\tr(\theta))^2 - \tr(\theta^2) \right] + \frac{2}{3} \tr(\theta^3) - \tr(\theta)\tr(\theta^2)
+\frac{1}{3} (\tr(\theta))^3 + 2 \theta^0_a \theta^a_b \theta^b_0 - 2 \tr(\theta) \theta_a^0 \theta^a_0
\nonumber
\end{eqnarray}
and from now on set $\varphi_1=\tilde{\varphi}_1$, $\varphi_0=\tilde{\varphi}_0$ so as to have consistent notation. 
Thus the mass term is of the form 
\begin{eqnarray}\label{generalmassterm}
{\cal L}_m(\varphi_0,\cdots,\varphi_3)
\end{eqnarray}
where ${\cal L}_m$ is any real valued (differentiable) function on $\mathbb R^4$. 
In terms of the new lapse and shift functions $\cal N$ and 
${\cal N}^a$ (see \ref{new lapse})),
a long but straightforward calculation leads to 
\begin{eqnarray}\label{generalbasisexpression}
\varphi_n = \alpha_n \, {\cal N} + {\cal N}^j (M_n)_j^i \chi_i + V_n
\end{eqnarray}
where $\alpha_n(E)$ and $V_n(E)$, as well as the 3-dimensional matrices $M_n(E)$, are independent of $\cal N$ and ${\cal N}^a$, and their explicit form is given in Appendix \ref{invariantfunctionsappendix}.

Finally we note that imposing that Minkowski metric is a solution, see (\ref{simp}), is equivalent to
\begin{eqnarray}\label{Minkowski metric}
\chi=0 \;\;\;,\;\;\; N^a=0 \;\;\;,\;\;\; N=1 \;\;\; \text{and} \;\;\; \theta=\mathbb I.
\end{eqnarray}

\section{Translation invariant fields: constraint analysis}
\label{section4}

In this section, we study the effect of introducing of a mass term in the canonical analysis of gravity. Thus we  consider the general action
\begin{eqnarray}\label{generalactionSm}
S[\theta_{(4)},\omega] = S_{grav}[\theta_{(4)},\omega] + S_m[\theta_{(4)}]  \;\;\;\; \text{with} \;\;\; S_m[\theta_{(4)}] =\int d^4x \, {\cal L}_m.
\end{eqnarray}
where $S_{grav}$ is the pure gravity action (\ref{puregravity}) and ${\cal L}_m$ is the general mass term (\ref{generalmassterm}).
In fact, we treat in detail the simpler case in which  all the fields depend only on the time variable $t$ --- namely we impose translation invariance.
The constraints analysis studied here will enable us to show, in section \ref{section5}, that the requirement that the theory has no ghost\footnote{with extra necessary assumptions to be detailed later.} imposes that $S_m$ is necessarily the standard dRGT form.  This is equivalent to saying that ${\cal L}_m$ must be linear in $N$ for there to be no ghost.

We note that in some analyses of massive gravity, e.g.~\cite{Hinterbichler:2012cn}, the vector field $\chi$ is fixed to zero as a partial gauge fixing (the so-called time gauge), but we will not do this here. Indeed, below we show that --- due to the form of the mass terms (\ref{generalmassterm}) --- $\chi$ will  vanish dynamically when translation invariance is imposed.

Before proceeding, notice that the assumption of time-dependence greatly simplifies the constraints in pure gravity (see section \ref{section2}).   Indeed the gravitational  vectorial constraint $H_a^{grav}$ (\ref{Ha}) no longer contains any derivatives, and it is straightforward to see that it is simply a linear combination of the constraint $S$ (\ref{S}) and $U$ (\ref{U}):
\begin{eqnarray}
H_a^{grav} & = & E^b\cdot F_{ba} + E^b \times \chi \cdot G_{ba}  \nonumber\\
& = & E^b \cdot (\omega_a \times A_b - \omega_b \times A_a) + E^b\times \chi \cdot (A_b \times A_a - \omega_b\times \omega_a)  \nonumber\\
& = & -\omega_a \cdot (E^b \times A_b - (\chi \times E^b)\times \omega_b) - A_a \cdot (E^b \times \omega_b + (\chi \times E^b)\times A_b)  \nonumber\\
& = & -\omega_a \cdot S - A_a \cdot U \,.
\label{salade}
\end{eqnarray}
Thus there are no more `vectorial constraints' (independent from the others), as expected in a homogeneous theory.

\subsection{Translation invariant fields in massive gravity: $\chi=0$}

We now consider the full action for massive gravity given in (\ref{generalactionSm}), which for translation invariant fields and on using
(\ref{salade}) becomes
\begin{eqnarray}\label{action invariance}
S & = & \int dt  [ E^a \cdot \partial_0 A_a - (\chi \times E^a) \cdot \partial_0\omega_a \nonumber \\ 
 & + & (A_0 -{\cal N}^a A_a)\cdot  U +  (\omega_0 - {\cal N}^a\omega_a)  \cdot S +\frac{\cal N}{2\vert E \vert^{1/2}} H^{grav}  + {\cal L}_m ].
\end{eqnarray}
Varying with respect to the Lagrange multipliers $\omega_0$ and $A_0$, and also the modified shift vector ${\cal N}^a$
yields the three primary constraints
\begin{eqnarray}\label{constrainttochi}
\frac{\delta S}{\delta A_0^i} = U_i \simeq 0 \;\;\;,\;\;\;
\frac{\delta S}{\delta \omega_0^i}=S_i \simeq  0  \;\;\;,\;\;\;
\frac{\delta S}{\delta {\cal N}^a} = -A_a \cdot U - \omega_a \cdot S +  \frac{\partial {\cal L}_m}{\partial {\cal N}^a} \simeq 0\, ,
\end{eqnarray}
where $\simeq$ denotes the weak equality (namely equality on the constraints surface). 
From the expression of the basis functions $\varphi_i$ given in (\ref{generalbasisexpression}), it therefore follows that 
\begin{eqnarray}\label{generalM}
  \frac{\partial {\cal L}_m}{\partial {\cal N}^a} =   \frac{\partial {\cal L}_m}{\partial \varphi_n}  \frac{\partial \varphi_n}{\partial {\cal N}^a}=  M^a_i \chi^i \simeq 0 \
 \;\;\;\text{where} \;\;\; M=  \frac{\partial {\cal L}_m}{\partial \varphi_n} M_n,
\end{eqnarray}
and the matrix $M_n$ was defined in (\ref{generalbasisexpression}).
Thus there are two branches of solutions to (\ref{generalM}):
\begin{enumerate}
\item $\chi=0$,
\item $\chi \neq 0 $ with $\det{M} = 0$.
\end{enumerate}
In the following we focus solely on the first branch which is guaranteed to be connected to Minkowski solution (\ref{Minkowski metric}).  From  the Hamiltonian point of view, this means not only that the initial conditions are such that $M$ is invertible, but also that the dynamics implies that this condition still holds for all times. This is not obviously the case, as has been discussed elsewhere  \cite{ms2}.
(In the second case, when $M$ is not invertible, the vector $\chi$ does not necessary vanish, so the classical theory does not have a Minkowski solution.)
In this first branch, the set of (9 scalar) constraints (\ref{constrainttochi}) is completely equivalent to the new set of (9 scalar) constraints
\begin{eqnarray}
U_i \simeq 0 \;\;\;,\;\;\; S_i \simeq 0 \;\;\;,\;\;\; \chi_i=0 \;\;\; \text{for} \;\; i \in \{1,2,3\}.
\end{eqnarray}
Since $\chi=0$ it follows from (\ref{new lapse}) that the modified lapse and shift correspond to the original ones, i.e. ${\cal N}=N$ and ${\cal N}^a=N^a$.

Before setting $\chi=0$  in  (\ref{action invariance}), however, consider the Euler-Lagrange equations obtained by varying with respect to $\chi$, namely
\begin{eqnarray}
\frac{\partial {\cal L}_m}{\partial \chi} - E^a \times \partial_0 \omega_a =0 \quad \text{with} \quad \frac{\partial {\cal L}_m}{\partial \chi^i}= \frac{\partial {\cal L}_m}{\partial \varphi_n} \frac{\partial {\varphi_n}}{\partial \chi^i}= (M^{\rm T})_{ia}N^a. 
\nonumber
\end{eqnarray}
Since this equation involves a time derivative,  it is an equation of motion and cannot --- at first sight --- be treated as a constraint. 
However, as we will see in section \ref{section4}, the components $\omega_a$
of the spin-connection actually vanish (independently of the form of the mass term). Thus, the previous equation does in fact becomes a constraint which, following the same arguments as for $\chi$,
leads to 
\begin{eqnarray}\label{Na vanish}
{\cal N}^a=N^a=0 \,.
\end{eqnarray} 
The same result (\ref{Na vanish}) was obtained from the Lagrangian point of view in \cite{ms1,ms2}. 
Notice that the conditions $\chi=0$ and $N^a=0$ are directly related to the translation invariance. Such constraints would not hold  for the general (not invariant under translations) theory of massive gravity.  (Notice that it is not a  priori necessary to fix  $N^a$ to zero because we can reabsorb it in a redefinition of the Lagrange multipliers $A_0$ and $\omega_0$ as follows:
$$
A_0 \longmapsto   A_0 +N^a A_a \, ,\qquad \omega_0 \longmapsto \omega_0 + N^a \omega_a \,.
$$
In a second step only, when we prove the constraint $\omega_a \simeq 0$, we could set $N^a=0$.  But to simplify the analysis we set $N^a=0$ immediately.)

\subsection{Constraint analysis}
\label{section canonical}
Our starting point for the constraints analysis is thus (\ref{action invariance}) in which we set $N^a=0$ and $\chi=0$;
\begin{eqnarray}
S =  \int dt \,  \left( E^a \cdot \partial_0 A_a + A_0 \cdot  U +  \omega_0 \cdot S +\frac{N}{2\vert E \vert^{1/2}} H^{grav}  + {\cal L}_m^0 \right)
\label{again}
\end{eqnarray}
where ${\cal L}_m^0$ denotes the mass term ${\cal L}_m$ with $\chi=0$. Note that ${\cal L}_m^0$ depends only on $N$ and $E$. 

\subsubsection{Primary constraints and total Hamiltonian}

In order to make contact with \cite{ms1,ms2}, from now on we use matrix rather than vector notation.
While $E$, $A$ and $\omega$ are by definition $3\times 3$ matrices, the Lagrange multipliers $A_0$ and $\omega_0$ have only one  free index.  However, we can identify them with antisymmetric matrices through
 \begin{eqnarray}
 A_0^{ij}=\epsilon^{ijk}A_{0k} \;\;\; \text{and} \;\;\; \omega_0^{ij}=\epsilon^{ijk}\omega_{0k},
 \nonumber
 \end{eqnarray}
so that (\ref{again}) can be rewritten as
\begin{eqnarray}
S =  \int dt \,  \left( \tr(E\cdot \partial_0 A)  + \tr(A_0 E\omega + \omega_0 EA) +\frac{N}{2\vert E \vert^{1/2}} H^{grav}  + {\cal L}_m^0 \right)
\label{effective action 2}
\end{eqnarray}
where the gravitational Hamiltonian constraint given in (\ref{H}) becomes
\begin{eqnarray}
H^{grav} = (\tr(EA))^2 -(\tr(E\omega))^2 - \tr((EA)^2-(E\omega)^2) .
\label{Gg}
\end{eqnarray}
From (\ref{effective action 2}) it follows that the non-trivial Poisson brackets are
\begin{eqnarray}\label{Poisson bra1}
\{E^a_i,A_b^j\} = \delta^a_b \, \delta_i^j  \; = \; \{P^a_i,\omega_b^j\} \;\;\text{and} \;\; \{N,\pi_N\}=1
\end{eqnarray}
where we add the momenta $P^a_i$ conjugate to the variables $\omega_a^i$, as before, and the momentum $\pi_N$ conjugate to the lapse $N$. 
This is necessary because the action is a priori
 a non-linear function of $N$, meaning that $N$ cannot be considered as a simple Lagrange multiplier. 

Since there are no time derivatives of $\omega$ and $N$, and since $A_0$ and $\omega_0$  are dynamical
Lagrange multipliers, it follows that the primary constraints are given by
\begin{eqnarray}
 C_P(u)&\equiv&\tr(uP) \simeq 0 \label{Cpdef} \\
C_\omega(v)&\equiv&\tr(vE\omega) \simeq 0 \,,\label{Comegadef} \\
C_A(v)&\equiv&\tr(vEA) \simeq 0 \,, \label{CAdef}\\
\pi_N &\simeq& 0,
\label{CAhello}
\end{eqnarray} 
for any $3\times 3$ matrix $u$ and any antisymmetric matrix $v$.  
The total Hamiltonian $H_{tot}$ is thus given by
\begin{eqnarray}\label{Haminv1}
-H_{tot}= C_\omega(A_0) + C_A(\omega_0) + C_P(u_0) +\mu \pi_N +  \frac{N}{2\vert E \vert^{1/2}} H^{grav}  + {\cal L}_m^0 
\end{eqnarray}
where $u_0$ and $\mu$ play the r\^ole of a Lagrange multipliers, enforcing the constraints $C_P(u)  \simeq 0$ and $\pi_N\simeq 0$.
The total Hamiltonian allows one to define the dynamics, and the time derivative of any function $\phi$ is given by
\begin{eqnarray}
\dot{\phi}=\{H_{tot},\phi\}.
\label{timeevoln}
\end{eqnarray}
Contrary to the situation in pure gravity, notice that the total Hamiltonian does not vanish: $H^{grav}$ is no longer a constraint (and there is no reason that  ${\cal L}_m^0$ should vanish).

\subsubsection{Secondary constraints: vanishing of spacelike spin-connection components}
\label{vanishingofomega}

Following the Dirac algorithm, we now study the stability of the primary constraints under time evolution.

On using (\ref{Cpdef}), (\ref{Haminv1}) and (\ref{timeevoln}), we find that the time derivative of the constraint $C_P(u)$
is given by
\begin{eqnarray}
\dot{C}_P(u) = \tr(A_0 Eu) + \frac{N}{\vert E \vert^{1/2}} \left( \tr(E\omega Eu) - \tr(E\omega)\tr(Eu) \right)\,\simeq 0
\label{ok}
\end{eqnarray}
which is totally independent of the mass terms. Indeed, the analysis of the stability of $C_P(u)\simeq 0$ is exactly as in pure gravity in the time gauge (see e.g.~\cite{GeillerNoui} and references therein).  Thus, following the standard analysis, we can show that amongst these 9 equations in (\ref{ok}), only 6 are  secondary constraints, since the remainder fix the 3 Lagrange multipliers $A_0$. Indeed, in component form (\ref{ok})  becomes
\begin{eqnarray}
\dot{C}_{P}(u^a_i) =\epsilon_{ijk}A_0^jE^{ka} + \frac{N}{\vert E \vert^{1/2}} \left( E^{c}_i\omega_c^j E^{ja} - \tr(E\omega) E^{a}_i \right)\,\simeq 0
\nonumber
\end{eqnarray}
where $(u^a_i)$ denotes the canonical basis of the space of $3\times 3$ matrices, and we now project these 9 equations onto the three dimensional vector basis $(E^{1},E^{2},E^{3} )$ to obtain
\begin{eqnarray}
\Psi^{ab} \equiv \dot{C}_{P}(u^a_i) E^{ib} = \epsilon_{ijk}A_0^jE^{ka}E^{ib} + \frac{N}{\vert E \vert^{1/2}} \left( (E^b \cdot E^{c})( \omega_c \cdot E^{a}) - \tr(E\omega) E^{a}\cdot E^b \right)\,\simeq 0.
\nonumber
\end{eqnarray}
(This is obviously equivalent to $\dot{C}_P(u) \simeq 0$ due to the invertibility of the matrix $E$.) At this point, we separate $\Psi^{ab}$ into its symmetric and antisymmetric components
\begin{eqnarray}
\Psi^{(ab)} & = & \frac{N}{\vert E \vert^{1/2}} \left( (E^c \cdot E^{(b})( E^{a)}  \cdot  \omega_c) - \tr(E\omega) E^{a}\cdot E^b \right)\,\simeq 0 
\nonumber
\\
\Psi^{[ab]} & = & A_0\cdot (E^a\times E^b) + \frac{N}{\vert E \vert^{1/2}} (E^c \cdot E^{[b})( E^{a]}  \cdot  \omega_c) \simeq 0
\nonumber
\end{eqnarray}
and notice that $\Psi^{(ab)}\simeq 0$ are new (secondary) constraints  (because $N$ should not vanish) whereas $\Psi^{[ab]}\simeq 0$ are equations which fix the Lagrange multiplier $A_0$ in terms
of $N$, $\omega_a$ and $E^a$. The physical meaning  of these new constraints $\Psi^{(ab)}\simeq 0$ is immediate: they are components of the space-like torsionless equation for the connection $\omega$ with respect to
the spatial triads. They are the translation invariant version of (\ref{Psiconstraint}).
If we add to these 6 constraints the three primary ones $C_\omega(v) \simeq 0$, we obtain exactly the 9 components of the torsionless equation for $\omega$. As a consequence, we can replace
the constraints  $\Psi^{(ab)}\simeq 0$ and $C_\omega(v)\simeq 0$ by the torsionless equation for $\omega$ or equivalently by the identification of $\omega$ with the space-like Levi-Civita connection.
Furthermore,  the homogeneity together with $\chi=0$ implies that the space like Levi-Civita connection vanishes identically. Thus, we have the following equivalence between secondary constraints
\begin{eqnarray}
\Psi^{(ab)}\simeq 0  \;\;\;\;\; \text{and} \;\;\;\;  C_\omega(v)\simeq 0 \; \Longleftrightarrow \; \omega_a^i \simeq 0.
\nonumber
\end{eqnarray}
Moreover, the three remaining equations $\Psi^{[ab]}=0$ fix the Lagrange multiplier $A_0$ to zero, i.e. $A_0\simeq0$ (on the constraints surface), as a consequence of $\omega_a\simeq 0$.  
Finally, as $\omega$ and $P$ vanish, we can just eliminate them from the 
original action so that the total Hamiltonian (\ref{Haminv1}) simplifies to
\begin{eqnarray}\label{simpler Htot1}
-H_{tot}=  C_A(\omega_0)   + \mu \pi_N +\frac{N}{2\vert E \vert^{1/2}} H^{grav}  +  {\cal L}_m^0(N,E).
\end{eqnarray}
where, from (\ref{Gg}), $H^{grav}= (\tr(EA))^2  - \tr((EA)^2)$ since $\omega_a=0$.

\subsubsection{Remaining secondary constraints: effect of the mass terms and symmetry of $E$}
To complete the analysis, we need to determine the time evolution of $C_\omega$, $C_A(v)$ and $\pi_N$ in (\ref{Comegadef}), (\ref{Cpdef}), and (\ref{CAhello}) respectively.
The first, $\dot{C}_\omega$  generates no further constraint since $\dot{C}_\omega \simeq 0$  fixes partially\footnote{To be complete, note that $u_0$ has nine scalar components.
Time evolution of $C_\omega$ allows one to fix 3 components. The 6 remaining are fixed from the time evolution of $\Psi^{(ab)}$.} the Lagrange multipliers $u_0$. 
In the case of $C_A$,  on using (\ref{CAdef}), (\ref{timeevoln}) and (\ref{simpler Htot1}), 
\begin{eqnarray}
\dot{C}_A(v) 
& \simeq  &   \{C_A(\omega_0) ,C_A(v)\} + \frac{N}{2\vert E \vert^{1/2}} \{H^{grav},C_A(v)\}  +  \{{\cal L}_m^0,C_A(v)\} ,
\nonumber
\end{eqnarray}
where one can show that
\begin{eqnarray}
\{C_A(v),H^{grav}\} &=& 0 \, ,
\nonumber
\\
\label{poissonCA}
\{C_A(v_1),C_A(v_2)\} &=& C_A([v_1,v_2]) \simeq 0 \, ,
\\
 \{ E,C_A(v)\}&=& vE \, ,
 \label{ppq} \\
 \label{poissonpourri}
\{{\cal L}_m^0,C_A(v)\}&=& \frac{\partial {\cal L}_m^0}{\partial E_i^a} \, \{E_i^a,C_A(v) \}
\end{eqnarray}
where $[v_1,v_2]$ is the matrix commutator.  These results are easily interpreted. Indeed, in pure gravity (when there are no mass terms), the $C_A(v)$ generate  the local $SU(2)$ invariance (which is generically 
broken when there is a mass term).  For this reason, they satisfy the Poisson algebra (\ref{ppq}) which is the $\su(2)$ Lie algebra, and from (\ref{poissonpourri}) its action on the matrix $E$ is the expected left action
of $\su(2)$ on the triads. 
Collecting these results together, the time derivative of the constraint $C_A(v)$ reduces, on the constraints surface, to 
\begin{eqnarray}\label{symmetry}
\dot{C}_A(v) \simeq \tr(vEL) \;\;\; \text{with} \;\;\; L_a^i= \frac{\partial {\cal L}_m^0}{\partial E_i^a} \,.
\end{eqnarray}
Due to the general expression of the mass term ${\cal L}_m^0$, the matrix $L$ is necessarily\footnote{Due to the form of ${\cal L}_m^0$, $L$ is necessary a linear combination of (negative and positive) 
powers of $E$. As $E$ is 3 dimensional, it satisfies the relation (\ref{Cayley}) where $\theta$ is replaced by $E$ (this is a direct consequence of  the Cayley-Hamiltonian theorem).
Therefore, any linear combination of powers of $E$ reduces to a linear combination of $\mathbb I$, $E^{-1}$ and $E^{-2}$ only. } of the form
\begin{eqnarray}\label{xyz}
L= x {\mathbb{I}}+ y E^{-1} + z E^{-2}
\end{eqnarray}
where $x$, $y$ and $z$ are scalar functions of $E$ and $N$. As a consequence (recall that $v$ is anti-symmetric),  (\ref{symmetry}) reduces to 
\begin{eqnarray}
\dot{C}_A(v) =\tr\left( v (xE+zE^{-1})\right)  \simeq 0 \,.
\label{talvste}
\end{eqnarray}
We see that $E$ symmetric is a solution of this equation --- but it might not be the unique one. For  particular expressions of $x$ and $z$ one could find non symmetric solution to the previous
constraint. (For an analogous discussion in the Lagrangian framework, see \cite{dmz1}).  However, as we are working in the sector where the Minkowski space-time is a solution, we consider only the solution $E$ symmetric.
Thus, the stability of $C_A(v)$ under time evolution leads to the following new secondary constraint:
\begin{eqnarray}
{C}_E(v) \equiv  \tr(vE) \simeq 0 \;
\label{CEhello}
\end{eqnarray}
where $v$ is an antisymmetric matrix.   This condition is crucial to show the equivalence between the metric and vielbein formulation of massive gravity, e.g.~\cite{dmz1}.

The time evolution of the last primary constraint $\pi_N$ produces also a new (scalar) secondary constraint
\begin{eqnarray}\label{secondaryH}
K \equiv \dot{\pi}_N =  \frac{H^{grav}}{2\vert E \vert^{1/2}}  + \frac{\partial {\cal L}_m^0}{\partial N} \simeq 0.
\end{eqnarray}
There are no more secondary constraints in the theory.

\subsubsection{Absence of ghost implies mass terms must be linear in $N$}

We now finish the constraint analysis by computing the time evolution of the secondary constraints $K$ and $C_E(v)$. Their stability will lead to a fixation of Lagrange multipliers and not to tertiary constraints when ${\cal L}_m^0$
is non-linear in $N$. 

 Using (\ref{CEhello}), the time evolution of the constraint $C_E(v)$ is given by
\begin{eqnarray}
\dot{C}_E(v) & = & \{C_E(v),H_{tot}\} \simeq  \{C_A(\omega_0),C_E(v)\} + \frac{N}{2\vert E \vert^{1/2}} \{ H^{grav},C_E(v)\} \nonumber \\
   & \simeq &   -\tr(\omega_0 Ev) + \frac{N}{\vert E \vert^{1/2}}  \tr(EAEv) \simeq 0 
   \nonumber
\end{eqnarray}
thus leading to the new relation
\begin{eqnarray}\label{fixomega0}
 \tr(\omega_0 Ev)  \simeq \frac{N}{\vert E \vert^{1/2}}  \tr(EAEv) \;\;\;\; \text{for any antisymmetric matrix} \; v.
\end{eqnarray}
 Up to some conditions to be discussed below, this relation fixes the Lagrange multipliers $\omega_0$ in terms of 
the lapse $N$ and the dynamical variables $E$ and $A$.
To see this, we evaluate this relation in the basis $(\epsilon_i)$ of antisymmetric matrices defined by $(\epsilon_i)_{jk}=\epsilon_{ijk}$ to obtain
 \begin{eqnarray}
 \omega_0^{jk} E^{k\ell} \epsilon_{i\ell j}=(E - \tr(E))_{ij}\omega_0^j  =  \frac{N}{\vert E \vert^{1/2}}  \tr(EAE\epsilon_i)
 \nonumber
\end{eqnarray}
where we make use of $\omega_0^{ij}=\epsilon^{ijk}\omega_{0k}$.
Thus $\omega_0$ is fixed provided the matrix $E-\tr(E)\mathbb{I}$ is invertible. 
When this condition is not satisfied,  Minkowski will not be a solution of the theory. Since we assume it is (see (\ref{Minkowski metric})), 
then we obtain
\begin{eqnarray}\label{omega0}
\omega_0 = \frac{N}{\vert E \vert^{1/2}} [(E - \tr(E))^{-1}]^{ij}  \tr(EAE\epsilon_j) .
\end{eqnarray}

Finally, the time evolution of the last secondary constraint $K$ (\ref{secondaryH}) is given by 
\begin{eqnarray}\label{secondderiv}
\dot{K} 
 & \simeq  & -\mu \frac{\partial^2 {\cal L}_m^0}{\partial N^2} + \Upsilon(N,E,A) \nonumber
\end{eqnarray}
where  
\begin{eqnarray}
\Upsilon \equiv  \{C_A(\omega_0) + \frac{N}{2\vert E \vert^{1/2}} H^{grav}+ {\cal L}_m^0 ,  \frac{H^{grav}}{2\vert E \vert^{1/2}} + \frac{\partial {\cal L}_m^0}{\partial N} \} 
\end{eqnarray}
whose explicit form is not needed in this section. If the mass term is non-linear in the lapse,
$$\frac{\partial^2 {\cal L}_m^0}{\partial N^2} \neq 0$$ 
equation (\ref{secondderiv}) is not a tertiary constraint but fixes the remaining Lagrange multiplier $\mu$ of the theory.
More precisely, the two equations (\ref{fixomega0}) and (\ref{secondderiv}) together are a system of four equations which allow one to fix the (four components of the) Lagrange multipliers $\mu$ and $\omega_0$
in terms of the dynamical variables in the theory. In other words, when ${\cal L}_m$ is assumed to be non-linear in the lapse, there is no  tertiary constraint  in the theory and the Dirac 
algorithm closes here.

To conclude, we compute the number of physical degrees of freedom in such a theory of massive gravity. First, we emphasize that all the constraints we found are necessarily second class. The reason is that
the mass term breaks the invariances of general relativity and there exist no new symmetries in the theory.  (This could be shown explicitly with a calculation of the Dirac matrix of the 
constraints --- however, we will do this in the following section for the usual dRGT mass terms.)
We started with the non-physical phase space which possesses $(9\times 2)+(9\times 2)+2=38$ (non physical) degrees  of freedom  (\ref{Poisson bra1}). Then, we found the following set of 26 second class constraints
\begin{eqnarray}
P^a \simeq 0 \;\;,\;\; \omega_a\simeq 0 \;\;,\;\; \pi_N\simeq 0 \;\;,\;\;
C_A(v)\simeq 0 \;\;,\;\; C_E(v)\simeq 0 \;\; \text{and}\;\;K\simeq 0
\nonumber
\end{eqnarray}
whose expressions have been given previously.
 This leads immediately to $38-26=12$ physical degrees of freedom in the phase space
of massive gravity. The ghost is present. The only hope to suppress the ghost is to consider a theory with a mass term linear in the lapse function $N$, which is the subject of the next section.

\section{Massive gravity with no ghost: Hamiltonian analysis}
\label{section5}

We now focus on a Lagrangian for the mass term which is linear in the lapse, or equivalently ${\cal L}_m(\varphi_0,\cdots,\varphi_3)$ which is linear in $(\varphi_{0},\cdots \varphi_{3})$, see (\ref{generalmassterm}).
 Thus we write
\begin{eqnarray}
{\cal L}_m[\theta_{(4)}]   =   {\cal L}_0+{\cal L}_1+ {\cal L}_2+{\cal L}_3 
\label{mass terms}
\end{eqnarray}
with 
\begin{eqnarray}
 {\cal L}_0 & = & \frac{\beta_0}{4!} \epsilon_{IJKL}   \theta^I \wedge \theta^J \wedge   \theta^K \wedge \theta^L \label{ll0} \\
 {\cal L}_1 & = & \frac{\beta_1}{3!}   \epsilon_{IJKL}  f^I \wedge \theta^J \wedge   \theta^K \wedge \theta^L  \label{ll1} \\
 {\cal L}_2 & = & \frac{\beta_2}{2!}  \epsilon_{IJKL}   f^I \wedge f^J \wedge   \theta^K \wedge \theta^L \label{ll2} \\
 {\cal L}_3 & = & \beta_3   \epsilon_{IJKL}   f^I \wedge f^J \wedge  f^K \wedge \theta^L \label{ll3} 
\end{eqnarray} 
The condition (\ref{nearly}) for Minkowski space to be a solution implies that the 4 parameters $\beta_{0,1,2,3}$ satisfy
\be
\beta_0 + 3\beta_1 + 6\beta_2+ 6\beta_3 = 0
\label{nearly2}
\ee
We will see that, in the Hamiltonian framework (and on using many of the results of the previous section)
these mass terms will lead generically to a theory of massive gravity with no ghost. 

\subsection{Hamiltonian form of dRGT massive gravity}

Initially, we do not consider a translation-invariant system and hence start by expressing (\ref{mass terms}) in its full Hamiltonian form. For clarity, we  compute each mass term  Lagrangian density ${\cal L}_n$ separately, and a long but straightforward calculation leads (see Appendix \ref{appendix2}) to
\begin{eqnarray}
{\cal L}_0       &=& \beta_0 \vert E \vert^{1/2} \left( N - N^a \theta_a \cdot \chi\right) = \beta_0 \vert E \vert^{1/2} (1-\chi^2) {\cal N} , \nonumber \\
{\cal L}_1 & = & \beta_1 \left( \vert E \vert^{1/2} + {\cal N}[(1-\chi^2)\text{tr}(E) +\chi \cdot (E \chi)] + \vert E \vert^{1/2}{\cal N}^a\chi_a  \right) ,\nonumber \\
{\cal L}_2 & = &  2\beta_2 \left( N \tr(\theta) + \tr(E) - N^a(\theta^2)^i_a \chi_i \right) , \nonumber \\
{\cal L}_3 & = & 6\beta_3(N + \tr(\theta)) = 6 \beta_3 \left( \vert E \vert^{1/2} \tr(E^{-1}) + {\cal N} + \vert E \vert^{1/2} {\cal N}^a (E^{-1}\chi)_a\right).
\nonumber
\end{eqnarray}
Before going further, notice that  ${\cal L}_1$ involve only the matrix $E$; ${\cal L}_3$ involves only the inverse matrix $E^{-1}$; and on the contrary 
${\cal L}_2$ involves the matrix $E$ and its inverse. Massive gravity with only the $\beta_2$ mass-term is more subtle and involved to analyze than when only $\beta_1$ or only $\beta_3$ are present (see e.g.~\cite{dmz1}).
As we will see in  following sections, we will recover these subtleties in the canonical analysis.
The total Lagrangian density (\ref{mass terms}) is then given by 
\begin{eqnarray}
{\cal L}_m  = V +  \frac{\cal N}{2\vert E\vert^{1/2}} H^m + {\cal N}^a H_a^{m}
\nonumber
\end{eqnarray}
where 
\begin{eqnarray}
 V & = &  \beta_1 \vert E \vert^{1/2} + 2\beta_2 \tr(E) + 6 \beta_3 \vert E \vert^{1/2} \tr(E^{-1}) \label{V}, \\
     {H^m}          & = &   2\beta_0 \vert E \vert(1-\chi^2) +  2\beta_1\vert E \vert^{1/2}[(1-\chi^2)\tr(E) + \chi \cdot (E\chi)]    \nonumber\\
        &&+\; \; 4\beta_2 \vert E \vert[\tr(E^{-1}) -\chi \cdot (E^{-1}\chi)]   + 12 \beta_3 \vert E \vert^{1/2} \label{Hm} ,\\
        H_a^{m}          & = & \beta_1 \vert E \vert^{1/2} \chi_a + 2\beta_2[(E\chi)_a - \tr(E) \chi_a] + 6\beta_3 \vert E \vert^{1/2} (E^{-1} \chi)_a.  \label{Ham}
\end{eqnarray}
(Below we will see that the physical Hamiltonian 
of massive gravity, once the second class are resolved, is given precisely by $-V$.)
Thus the total massive gravity action takes the Hamiltonian form
\begin{eqnarray}\label{action ham}
S & = & \int dt \int d^3 x \,  [ E^a \cdot \partial_0 A_a - (\chi \times E^a) \cdot \partial_0\omega_a \nonumber \\ 
    & + & A_0 \cdot  U +  \omega_0 \cdot S +\frac{\cal N}{2\vert E \vert^{1/2}} (H^{grav} + H^{m}) + {\cal N}^a (H_a^{grav} + H_a^{m}) +  V ]
\end{eqnarray}
where the different functions involved are defined in (\ref{U}),(\ref{S}),(\ref{H}),(\ref{Ha}),(\ref{V}),(\ref{Hm}),(\ref{Ham}).

\subsection{Translation invariance: constraints and symmetry of $E$ revisited}

From now on, and for the remainder of this paper, we consider only translation invariant systems and thus all fields only depend on $t$.
 Then many of the results of the previous section hold: in particular, we can set $\chi=0$ and $N^a=0$ directly in (\ref{action ham}), which we can also write in matrix form as before to obtain
\begin{eqnarray} \label{effective action}
S =  \int dt \,  \left( \tr(E\cdot \partial_0 A)  + \tr(A_0 E\omega + \omega_0 EA) +\frac{N}{2\vert E \vert^{1/2}} (H^{grav} + H^{m})  +  V \right),
\end{eqnarray}
where the two terms of the Hamiltonian constraints are given by
\begin{eqnarray}
H^{grav}& =& \big(\tr(EA)\big)^2 -\big(\tr(E\omega)\big)^2 - \tr\big((EA)^2-(E\omega)^2\big), \label{Hgravy} \\
H^{m} &  = &  2\vert E \vert^{1/2}\left( \beta_1\tr(E) + \beta_0\vert E\vert^{1/2} + 4\beta_2 \vert E \vert^{1/2} \tr(E^{-1}) +6\beta_3 \right).\label{Hmgravy}
\end{eqnarray}

From action (\ref{effective action}), the non-trivial Poisson brackets are
\begin{eqnarray}\label{Poisson bra}
\{E^a_i,A_b^j\} = \delta^a_b \, \delta_i^j  \; = \; \{P^a_i,\omega_b^j\}
\end{eqnarray}
where we add the momenta $P^a_i$ conjugate to the variables $\omega_a^i$. However, contrary to the case studied in the previous section, the Lagrangian density is now linear in $N$
so that there is no need to introduce a conjugate momentum to the lapse; in other words, the lapse can be considered directly as a Lagrange multiplier 
exactly as  $\omega_0$ and $A_0$ which come with no time derivatives. 
These Lagrange multipliers, together with the fact that $\omega$ is non-dynamical impose the same 3 primary constraints as in (\ref{Cpdef})-(\ref{CAdef}), namely
\begin{eqnarray}
 C_P(u)=\tr(uP) \simeq 0 \, , \qquad
C_\omega(v)=\tr(vE\omega) \simeq 0 \, \qquad
C_A(v)=\tr(vEA) \simeq 0 \,
\end{eqnarray}
for any $3\times 3$ matrix $u$ and any antisymmetric matrix $v$.
The constraint $\dot{\pi}_N\simeq 0$ in (\ref{secondaryH}) can be rewritten as 
\begin{eqnarray}
H& \equiv &H^{grav}+H^m \simeq 0,
\label{Hope}
\end{eqnarray} 
(where $H=2|E|^{1/2}K$). There are no more primary constraints, and the total Hamiltonian is given by
\begin{eqnarray}\label{Haminv}
-H_{tot}= C_\omega(A_0) + C_A(\omega_0) + C_P(u_0) +\frac{N}{2\vert E \vert^{1/2}} H  +  V 
\end{eqnarray}
where $u_0$ plays the role of a Lagrange multiplier.  
Notice that, contrary the situation in pure gravity, and due the presence of the potential $V$, the total Hamiltonian does not vanish.

The study of the stability of the constraint $C_P(u)$ is totally independent of the mass terms, and thus the analysis done in subsection \ref{vanishingofomega} applies directly: thus  $\omega$ and $P$ vanish and we can just eliminate them from the 
original action. Therefore, the total Hamiltonian (\ref{Haminv}) simplifies to
\begin{eqnarray}\label{simpler Htot}
-H_{tot}=  C_A(\omega_0)  +\frac{N}{2\vert E \vert^{1/2}} H  +  V 
\end{eqnarray}
where $H$ is defined in (\ref{Hope}), $V$ is as given in (\ref{V}), and from (\ref{Hgravy}) 
\be
H^{grav}= (\tr(EA))^2  - \tr((EA)^2).
\label{Hgravy2}
\ee

The study of the time evolution of $C_A(v)$ is also very similar to the general case: it is given by (\ref{talvste}), namely
\begin{eqnarray}
\dot{C}_A(v) \simeq  \tr\left(v(xE+zE^{-1})\right)   \simeq 0 
\nonumber
\end{eqnarray}
where on using
\begin{eqnarray}
\{C_A(v),H^{m}\} & = & -2\beta_1 \vert E \vert^{1/2} \tr(vE) + 4 \beta_2 \vert E \vert \tr(vE^{-1})\nonumber \\
\{C_A(v),V\} & = & -2\beta_2 \tr(vE) + 6\beta_3 \tr(vE^{-1})\,
\nonumber
\end{eqnarray}
we find that
\be
 x=  \beta_1N + 2\beta_2 \;\;\;\text{and}\;\;\; z= -N\vert E \vert^{1/2} \beta_2 - 6\beta_3.
 \nonumber
 \ee
Thus $\dot{C}_A(v)\simeq 0$ implies that $xE+zE^{-1}$ is a (weakly) symmetric matrix. 
This constraint is solved by a matrix $E$ which is symmetric\footnote{When only $\beta_1$ or $\beta_3$ are non-vanishing, the condition $E$ symmetric is equivalent to the constraint. 
When only $\beta_2 \neq 0$,  the constraint admits other solutions than $E$ 
symmetric. When at least 2 parameters out of the 3 are non-vanishing,  $E$ symmetric may not be the unique solution of the constraint as well.} (though as discussed above there may exist other solutions (non symmetric $E$) but we do not consider them because we want the Minkowski space-time  to be a solution). 
Thus once again, we find the secondary constraint (\ref{CEhello})   
\begin{eqnarray}\label{symmE}
C_E(v) = \tr(vE) =0 \;\;\;\; \text{for any} \;\;\; v^{\rm T}=-v
\end{eqnarray}
as in the previous section. 

\subsection{A new scalar constraint removes the ghost }
Now we study the evolution of the remaining primary constraint $H$ given in (\ref{Hope}), namely
\begin{eqnarray}
\dot{H}&=&\{C_A(\omega_0) + \frac{N}{2\vert E \vert^{1/2}} H + V,C_A(v)\} 
\nonumber
\\
&\simeq &\frac{N}{2\vert E \vert^{1/2}} \{ C_A(\omega_0),H^{m}\} + \{V,H^{grav}\}.
\nonumber
\end{eqnarray}
Due to the secondary constraint (\ref{symmE}), the first Poisson bracket in the previous equation vanishes on the constraint surface because
\begin{eqnarray}
\{C_A(\omega_0),H^{m}\}  = -2\beta_1 \vert E \vert^{1/2} \tr(\omega_0E) + 4 \beta_2 \vert E \vert \tr(\omega_0 E^{-1}) \simeq 0\;.
\nonumber
\end{eqnarray}
To evaluate the second Poisson bracket, it is useful to notice from (\ref{Hgravy2}) that
\begin{eqnarray}
\{H^{grav},E\} = 2 (EAE -\tr(EA)E) & \Longrightarrow &\{H^{grav},\vert E \vert \} = -4\tr(EA) \vert E \vert \\
&  \text{and}  & \{H^{grav},E^{-1}\} = -2(A - \tr(EA)E^{-1})
\nonumber
\end{eqnarray}
from which we obtain (after a long but straightforward calculation)
\begin{eqnarray}
\dot{H} \simeq  \{V,H^{grav}\} \simeq 2 \left( \beta_1 \vert E \vert^{1/2} \tr(EA) - 2 \beta_2 \tr(EAE) + 6\beta_3 \vert E \vert^{1/2} \tr(A)\right).
\nonumber
\end{eqnarray}
Hence the time stability of the constraint $H \simeq 0$ implies a new secondary constraint
\begin{eqnarray}
\Psi \equiv  \beta_1 \vert E \vert^{1/2} \tr(EA) - 2 \beta_2 \tr(EAE) + 6\beta_3 \vert E \vert^{1/2} \tr(A)  \simeq 0\,.
\label{Psidef}
\end{eqnarray}

This strongly contrasts with the previous section. Before looking for  tertiary constraints, it is useful to make a short summary at this point. The phase space reduces now to the pair of dynamical variables $(E^a_i,A_b^i)$ which are conjugate according
to the Poisson bracket (\ref{Poisson bra}). The variables $\omega$ and $P$ were shown 
to vanish (essentially due to translation invariance) together with the variables $\chi$ and $N^a$. The remaining dynamical variables are subject to 4 primary constraints
\begin{eqnarray}
C_A(v)&=&\tr(vEA) \simeq 0 \nonumber \\
 H&=&H^{grav} + H^m \simeq 0 \label{CEV}
 \end{eqnarray}
 where  $v$ is any anti-symmetric matrix, $H^m$ is given in (\ref{Hmgravy}) and $H^{grav}$  in (\ref{Hgravy2}) .
To these primary constraints, we add 4 secondary ones obtained from the requirement of the stability under time evolution of the previous ones, namely
\begin{eqnarray}
C_E(v)= \tr(vE)&\simeq& 0 \nonumber \\
  \Psi  & \simeq &0
  \label{secondary constraints}
\end{eqnarray}
where $\Psi$ was defined in (\ref{Psidef}).
Thus we have obtained 8 constraints starting with 
18 non-physical degrees of freedom. As the system possesses no local symmetries (the mass terms break the local symmetries of gravity), it is clear that these 8 constraints are second class.
Therefore, we do not expect to get more (tertiary) constraints, but rather to finish the Dirac analysis with 10 physical degrees of freedom as expected in a massive theory of gravity. In the following
subsection we verify that this is indeed the case. 

\subsection{Fixation of Lagrange multipliers}
\label{fixLag}
We now show that 
the stability under time evolution of the secondary constraints leads to a fixation of Lagrange multipliers and not to tertiary constraints. 

The study $\dot{C}_E(v)$ works exactly as in the previous section, and so (as before)
  $\omega_0$ is fixed in terms of the lapse function $N$
and the dynamical variables --- provided the matrix $E-\tr(E)$ is invertible. In that case
\begin{eqnarray}\label{omega0}
\omega_0 = N \tilde{\omega}_0 \;\;\; \text{with} \;\;\; \tilde{\omega}_0^i= \frac{N}{\vert E \vert^{1/2}} [(E - \tr(E))^{-1}]^{ij}  \tr(EAE\epsilon_j).
\end{eqnarray}
Time evolution of the last constraint $\Psi$ defined in (\ref{Psidef}) is given by
\begin{eqnarray}
\dot{\Psi} & = & \{C_A(\omega_0) + \frac{N}{2\vert E \vert^{1/2}} H+ V,\Psi\} 
\nonumber
\\
   & \simeq&  \frac{N}{2\vert E \vert^{1/2}} \{H,\Psi\} + N \{C_A(\tilde{\omega}_0) ,\Psi \} + \{V,\Psi\} \simeq 0,
    \nonumber
\end{eqnarray}
which fixes the last Lagrange multiplier, the lapse function $N$, provided
\begin{eqnarray}
  \{H,\Psi\} + 2\vert E \vert^{1/2} \{C_A(\tilde{\omega}_0) ,\Psi \} \neq 0.
  \nonumber
\end{eqnarray}
(In the next section we will see that this requirement is necessary for the Dirac matrix to be invertible.) Assuming it holds, then
\begin{eqnarray}
N \simeq 2\vert E \vert^{1/2} \frac{ \{\Psi,V\} }{\{H,\Psi\} + 2\vert E \vert^{1/2} \{C_A(\tilde{\omega}_0) ,\Psi \} }
\label{lapsypoos}
\end{eqnarray}
so that the Dirac algorithm closes.  Note that the lapse $N$ can be obtained explicitly by computing the Poisson brackets in the denominator of (\ref{lapsypoos}), but in general this leads to a complicated (and not very useful) expression, which is why  we do not give it here. 

However, as an example, and in order to check our calculations, let us compute $N$ explicitly when $\beta_2=\beta_3=0$ and thus, from (\ref{nearly2}), $\beta_0=-3\beta_1$: this is precisely the case studied in \cite{ms1}. 
Then the constraints simplify and, in particular $\{C_A(u) ,\Psi \} = 0$ for any antisymmetric matrix $u$, which implies that
\begin{eqnarray}\label{lapse expression}
N & \simeq & 2\vert E \vert^{1/2} \frac{ \{\Psi,V\} }{\{H,\Psi\} } \simeq \frac{ 2 \beta_1 \vert E \vert^{1/2}  \{\tr(EA),\vert E \vert^{1/2}\} }{\{-\tr((EA)^2) +2\vert E \vert^{1/2}\left( \beta_1\tr(E) + \beta_0\vert E\vert^{1/2}\right) ,\tr(EA)\} } \nonumber\\
& \simeq  & \frac{3}{18 - 5 \tr(\pi)}.
\end{eqnarray}
Here, in the last line, we have introduced the notation  $\pi=\vert E \vert^{-1/2}E$  in order to compare with \cite{ms1}.  (The calculations required to obtain the last line explained in more detail in section \ref{section:help}.)
This expression of the lapse is identical to equation (53) found in \cite{ms1} with $D=4$ and the redefinition\footnote{In \cite{ms1,ms2}, the lapse was defined by $-{N}^{-1}=\theta^{00}=\theta^{0}_0 f^{00}=-\theta^0_0$
whereas in this article we have defined $\theta_0^0=N$.} $N \rightarrow 1/N$.

\section{Time evolution of the dynamical variables}
\label{section6}

The constraint analysis presented above is useful in so far as it gives a formal definition of the physical phase space and, as an immediate consequence, the number of physical degrees of freedom in the theory.  However, in order to compute the equations of motion and (if possible) determine their solutions, we need to solve the second class constraints.  

To do so, we proceed via the calculation of the Dirac matrix of second class constraints \cite{Henneaux:1992ig}.  If we denote the 8 second class constraints by $S_i$ ($i=1,\ldots 8$), then recall that the Dirac matrix is defined by
\be
\Delta_{i j} = \{ S_i,S_j \}
\label{Diracmatrix}
\ee
from which the Dirac bracket for two phase space variables $f$ and $g$ is given by
\be
 \{ f,g \}_D =  \{ f,g \} -  \{ f,S_i \} (\Delta^{-1})^{ij} \{ S_j, g \}.
\label{Diracbracket}
 \ee
Thus we need to calculate and invert the Dirac matrix (subsection \ref{subsec6a}) in order to determine the equations of motion (subsection \ref{subsec6b}).

\subsection{Dirac matrix}
\label{subsec6a}

The 8 second class constraints are $(C_E(v), C_A(v), \Psi,H)$.  However,  it will be useful to redefine these slightly in order to calculate the Dirac matrix.  To do so, introduce the following (matrix) variables
\begin{eqnarray}
\pi = \vert E \vert^{-1/2} E \;\;\;\; \text{and} \;\;\;\; \Omega= \pi A
\nonumber
\end{eqnarray}
whose Poisson brackets (\ref{Poisson bra}) satisfy (see appendix \ref{Poisson OmegaPi})
\begin{eqnarray}
 \{\pi_{ij},\pi_{k\ell}\} &=& 0 
\nonumber
\\
\{\pi_{ij},\Omega_{k\ell}\} &=& \vert \pi \vert (\delta_{i\ell} \pi_{kj} - \frac{1}{2} \delta_{k\ell} \pi_{ij}) \;,
\nonumber\\
\{ \Omega_{ij},\Omega_{k\ell}\} &=& \vert \pi \vert (\delta_{i\ell} \Omega_{kj} - \delta_{jk} \Omega_{i\ell} -\frac{1}{2} \delta_{k\ell} \Omega_{ij} + \frac{1}{2} \delta_{ij} \Omega_{k\ell} ).
\nonumber
\end{eqnarray}
Then, we work with the following equivalent set of second class constraints 
\begin{eqnarray}
C_m &\equiv & \epsilon_{ijm} \pi_{ij} \; \simeq \; 0 \nonumber \\
 D_m &\equiv & \epsilon_{ijm} \Omega_{ij} \; \simeq \; 0  \nonumber
 \\
 \tilde{\Psi} & \equiv & \vert \pi \vert \Psi  =   \vert \pi \vert^{-1} \left( \beta_1 \tr(\Omega) - 2 \beta_2 \tr(\Omega \pi)\right) + 6\beta_3 \tr(\Omega \pi^{-1}) \simeq 0 \;,
 \label{Psidef2}
\\
\tilde{H} &\equiv& \vert \pi \vert^2 H = (\tr\Omega)^2 -\tr(\Omega^2) + 2  \left( \beta_0 + \beta_1 \tr(\pi) + 4\beta_2 \vert \pi \vert \tr(\pi^{-1}) + 6 \beta_3 \vert \pi \vert \right) \simeq 0.
\label{Hdef2}
\end{eqnarray}
The first two constraints replace respectively  $C_E(v)\simeq 0$ (\ref{CEV}) and $C_A(v)\simeq 0$ (\ref{secondary constraints}), and they implement the fact that $\pi$ and $\Omega$ are symmetric matrices.

%
%
%

We order the 8 second class constraints $S_j$ in the following way:
\be
S_{1,2,3} = C_{1,2,3} , \qquad S_{4,5,6} = D_{1,2,3}, \qquad S_7=\Psi, \qquad S_8 = H
\nonumber
\ee
where, for notational simplicity, we now drop the tilde's on $H$ and $\Psi$.
A long but straightforward calculation (the details are given in the appendix \ref{appendix Dirac}) then shows that the Dirac matrix (\ref{Diracmatrix}) takes the block matrix form
\begin{eqnarray}
\Delta=\left(\begin{array}{c|c|c|c}
0& {A} & 0 & -x\\
\hline
{-A}& 0 & -y & 0\\
\hline
0 & y^{\rm T} & 0 & a\\
\hline
x^{\rm T} & 0 & -a & 0 \\
\end{array}\right)
\label{difac}
\end{eqnarray}
where $A$ is a 3 dimensional matrix; $x$ and $y$ are 3 dimensional vectors; and $a$ is a scalar. These are given by
\begin{eqnarray}
A_{mn}&=& \{C_m,D_n\} \simeq \vert \pi \vert (\pi - \text{tr}(\pi))_{mn} \nonumber\\
x_m &= & \{ H,C_m\} \simeq 2 \vert \pi \vert \epsilon_{ijm} (\Omega \pi)_{ij}\nonumber\\
y_m &= & \{ \Psi,D_m\} \simeq  -2 \epsilon_{ijm}\left(\beta_2 (\Omega \pi)_{ij} + 3 \beta_3 \vert \pi \vert (\pi^{-1} \Omega)_{ij}\right),\label{yydef}  \\
a&=& \{\Psi,H\}
\nonumber
\end{eqnarray}
where the explicit expression for $a$ is not required in the general case (see below). These are the only (weakly) non-vanishing Poisson brackets between the
constraints.

\subsubsection{Inverting the Dirac matrix}

In order to compute the Dirac bracket (\ref{Diracbracket}), we need to invert the Dirac matrix $\Delta$. To do so we introduce an orthonormal basis $(e_m,f_m,e_4,f_4)$ with $i\in \{1,2,3\}$ in the 8-dimensional vector space of constraints, so that the matrix elements of $\Delta$ are given by
%
%
\begin{eqnarray*}
&&\langle e_m \vert \Delta \vert f_n \rangle= A_{mn} = - \langle f_m \vert \Delta \vert e_n \rangle \;\;\;\;\;,\;\;\;\;
\langle f_4 \vert \Delta \vert e_m \rangle= x_{m} = - \langle e_m \vert \Delta \vert f_4 \rangle \\
&&\langle e_4 \vert \Delta \vert f_m \rangle= y_{m} = - \langle f_m \vert \Delta \vert e_4 \rangle \;\;\;\;\;\;\;\;,\;\;\;\;
\langle e_4 \vert \Delta \vert f_4 \rangle= a = - \langle f_4 \vert \Delta \vert e_4 \rangle.
\end{eqnarray*}
Hence
\begin{eqnarray}
\Delta = M_{\mu\nu} \, (\vert e_\mu \rangle \langle f_\nu \vert - \vert f_\nu \rangle \langle e_\mu \vert )  \;\;\;\; \text{with} \;\;\;
M=\left(\begin{array}{c|c}
A & -x \\
\hline
y^{\rm T} & a \\
\end{array}\right)
\nonumber
\end{eqnarray}
where now the index $\mu$ runs in $\{1,2,3,4\}$ so that $M$ is a 4-dimensional matrix. Therefore, the problem of inverting the 8-dimensional matrix $\Delta$ reduces to 
the inversion of the 4-dimensional matrix $M$ since
\begin{eqnarray}
\Delta^{-1} =  -M^{-1}_{\nu\mu}  (\vert e_\mu \rangle \langle f_\nu \vert - \vert f_\nu \rangle \langle e_\mu \vert ) \;.
\nonumber
\end{eqnarray}
Thus $M$ must be invertible, and when this is the case, a straightforward calculation shows that
\begin{eqnarray}\label{inverseDirac}
M^{-1}=\left(\begin{array}{c|c}
B & -u \\
\hline
v^{\rm T} & b \\
\end{array}\right)  
\end{eqnarray}
where
\begin{eqnarray}
B&=&(A+\frac{1}{a} x\, y^{\rm T})^{-1} ,\;
\nonumber
\\ b &=& (a+y\cdot A^{-1}x)^{-1} \nonumber\\
  u&=&-bA^{-1}x\;,\nonumber\\
   v&=&-\frac{1}{a} B^{\rm T}y 
 \label{superuseful}
\end{eqnarray}
where $t$ denotes a transpose. Thus $M^{-1}$ exists only if the conditions 
\begin{eqnarray}\label{invertibility}
 \vert A \vert \neq 0 \;,\;\; a \neq 0\;,\;\; a+y\cdot A^{-1}x  \neq 0 \;
\end{eqnarray}
hold. These conditions ensure that the matrix $B$ defined above is invertible and therefore exists, since a direct calculation shows that its determinant is given by
\begin{eqnarray}
\vert B \vert^{-1}=\vert A+\frac{1}{a} x\, y^{\rm T} \vert = \frac{\vert A \vert}{a }  (a+y\cdot A^{-1}x).
\nonumber
\end{eqnarray}
We notice that the two first conditions in (\ref{invertibility}) are the same as the ones found in   subsection \ref{fixLag} in order to fix the Lagrange multipliers $\omega_0$
and $N$. It is indeed well known that the invertibility of the Dirac matrix is closely related to the fixation of  Lagrange multipliers.

Thus when the conditions (\ref{invertibility}) are satisfied, the inverse of the Dirac matrix (\ref{difac}) is given by
\begin{eqnarray}
\Delta^{-1}=
\left(\begin{array}{c|c|c|c}
0& {-B^{\rm T}} & 0 & -v\\
\hline
{B^{\rm T}}& 0 & -u & 0\\
\hline
0 & u^{\rm T} & 0 & -b\\
\hline
v^{\rm T} & 0 & b & 0 \\
\end{array}\right)
\nonumber
\end{eqnarray}
where the 3 dimensional matrix $B$, the 3 dimensional vectors $u$ and $v$, and the scalar $b$ have been defined in (\ref{inverseDirac}).

\subsubsection{General expression of the Dirac bracket}

We have now all the ingredient at hand to construct explicitly the Dirac bracket (\ref{Diracbracket}).  Notice that this definition insures that $\{f,S_i\}_D=0$, meaning that the second class constraints are implicitly imposed.
Also, one can show that the Dirac bracket defines a good Poisson structure: it is antisymmetric and satisfies the Jacobi identity. Due to the expression of the inverse Dirac matrix
$\Delta^{-1}$ in our case, the Dirac bracket is precisely given by
\begin{eqnarray}\label{Dirac bracket general}
\{f,g\}_D  =  \{f,g\} & - & B^{nm} (\{f,C_m\} \{D_n,g\} - \{f,D_m\} \{C_n,g\}) \nonumber \\
 &+ & v^m(\{f,H\}\{C_m,g\}-\{f,C_m\}\{H,g\}) \nonumber \\
 & + & u^m(\{f,\Psi\} \{D_m,g\}-\{f,D_m\} \{\Psi,g\} )  \nonumber \\
 &+ & b (\{f,H\}\{\Psi,g\}-\{f,\Psi\} \{H,g\}).
 \label{sds}
\end{eqnarray}
From this general definition, we will now extract the Dirac brackets between the fundamental variables $\pi$ and $\Omega$. This defines exactly and explicitly the physical phase space
of translation invariant time dependent massive gravity.

\subsection{Equations of motion}
\label{subsec6b}
From the Dirac bracket, we can write explicitly the Hamiltonian equations of motion:
\begin{eqnarray}\label{general eom}
\dot{\pi}_{ij} = \{ V,\pi_{ij}\}_D \;\;\;\; \text{and} \;\;\;\; \dot{\Omega}_{ij} = \{ V,\Omega_{ij}\}_D .
\end{eqnarray} 
As expected, on the constraints surface the total Hamiltonian reduces to the (minus the) potential $H_{tot}=-V$ (where $V$ is given in (\ref{V})).  Hence the function $-V$ plays the role of the Hamiltonian vector field for the dynamics on the 
constraints surface (with respect to the Dirac bracket), and it defines the energy of translation invariant but time dependent massive gravity. 

As $V$ depends only on $\pi$ (whatever the choice of $\beta_i$), then $\{V,\pi_{ij}\}=0=\{V,C_m\}$ and therefore  the equations of motion (\ref{general eom}) can be simplified and on using (\ref{sds}) take the general form
\begin{eqnarray}
\dot{\pi}_{ij}  & = &  u^m(\{V,\Psi\} \{D_m,\pi_{ij}\}-\{V,D_m\} \{\Psi,\pi_{ij}\} )  \nonumber \\
&&+ b (\{V,H\}\{\Psi,\pi_{ij}\}-\{V,\Psi\} \{H,\pi_{ij}\}) \label{ffpi} \\
\dot{\Omega}_{ij} & = & v^m \{V,H\}\{C_m,\Omega_{ij}\} + u^m(\{V,\Psi\} \{D_m,\Omega_{ij}\}-\{V,D_m\} \{\Psi,\Omega_{ij}\} )  \label{ffomega} \\
&&+ b (\{V,H\}\{\Psi,\Omega_{ij}\}-\{V,\Psi\} \{H,\Omega_{ij}\}). 
\nonumber
\end{eqnarray}
To obtain a more explicit  formula, it is necessary to compute the Poisson 
brackets between the fundamental variables ($\pi$,$\Omega$), the second class constraints and $V$. This has been done appendix \ref{appendix Dirac}. 
Using all these results, the equations of motion are obtained immediately, though the calculation is long and tedious.  As far as we can see the equations of motion do not appear to have any particular structure when all the mass terms are present, and hence
for that reason, we now illustrate the dynamics of the theory only in particular cases. 


\subsubsection{First order dynamical system}
\label{section:help}

To compare with the results of \cite{ms1}, we again consider  the case in which $\beta_2=\beta_3=0$, and $\beta_0=-3\beta_1$ with\footnote{In \cite{ms1}, the mass $m$ was defined by $\beta_1'=-2m^2$. As $\beta'_i=2\beta_i$ here, we obtain $\beta_1=-m^2$.} $\beta_1=-m^2$.

Then
the constraints $\Psi$, $H$ and $V$, given respectively in (\ref{Psidef2}), (\ref{Hdef2}) and (\ref{V}) simplify to
\begin{eqnarray}
\Psi=-m^2 \vert \pi \vert^{-1} \tr(\Omega) \;\;,\;\;\; H \simeq  -\tr(\Omega^2) + 2m^2(3-\tr(\pi)) \;\; \text{and} \;\;\; V =- m^2 \vert \pi \vert^{-1}
\nonumber
\end{eqnarray}
(where we have subtracted from $H$ the term proportional to $\tr(\Omega)$ since this vanishes on the constraints surface from the expression for $\Psi$).
Furthermore the expression of the Dirac bracket simplifies drastically because from (\ref{yydef}), $y=0$ and therefore from (\ref{superuseful})
\begin{eqnarray}
B=A^{-1} \;\;,\;\; b = \frac{1}{a} \;\;\text{and}\;\; u = -\frac{1}{a}A^{-1}x 
\nonumber
\end{eqnarray}
while (see Appendix \ref{appendix Dirac})
\be
 a = \{\Psi,H \} = m^4(18-5\tr(\pi)).
\nonumber
\ee
Thus the equations of motion (\ref{ffpi}) and (\ref{ffomega}) reduce to
\begin{eqnarray}
\dot{\pi}_{ij} &=& m^2 \vert \pi \vert^{-2}\{ \vert \pi \vert ,\pi_{ij}\}_D 
\nonumber\\
&=&m^2 \vert \pi \vert^{-2}
u^m \{ \vert \pi \vert , \Psi\} \{D_m,\pi_{ij}\} - \frac{1}{a} \{\vert \pi \vert , \Psi\} \{H,\pi_{ij}\} \label{ppp}
\\
 \dot{\Omega}_{ij} &=& m^2 \vert \pi \vert^{-2}\{ \vert \pi \vert,\Omega_{ij}\}_D
\nonumber \\
&=& \{ \vert \pi \vert,\Omega_{ij}\} + u^m\{ \vert \pi \vert , \Psi\} \{D_m,\Omega_{ij}\}  - \frac{1}{a} \{\vert \pi \vert , \Psi\} \{H,\Omega_{ij}\} \label{ooo}.
\end{eqnarray}
where all the Poisson brackets involved in formulae (\ref{ppp}) and (\ref{ooo}) are given by (see Appendix \ref{appendix Dirac})
\begin{eqnarray}
&&\{ \vert \pi \vert ,\Psi \} =  \frac{3}{2} m^2 \vert \pi \vert \;,\;
\{D_m,\pi_{ij}\} = \vert \pi \vert \epsilon_{mik} \pi_{kj} \;,\;
\{H,\pi_{ij} \} = 2\vert \pi \vert (\Omega\pi)_{ij}\;,\;  \{ \vert \pi \vert , \Omega_{ij} \} = Ð\frac{1}{2}\vert \pi \vert^2 \delta_{ij} \nonumber  \\
&& \{D_m,\Omega_{ij} \}=\vert \pi \vert (\epsilon_{mjk}\Omega_{ki} + \epsilon_{mik}\Omega_{kj}) \;,\;
\{H,\Omega_{ij} \}=m^2\vert \pi \vert ((6-\tr(\pi)) \delta_{ij} - 2\pi_{ij}).
\nonumber
\end{eqnarray}
As a consequence, a direct calculation leads to the following equations of motion for $\pi$ and $\Omega$
\begin{eqnarray}
\dot{\pi}_{ij} & = & -N \left(\frac{1}{2} (A^{-1}x)_m \epsilon_{mik} \pi_{kj} + (\Omega\pi)_{ij}\right) \label{pidot}\\
\dot{\Omega}_{ij} & = & N\left( m^2(\pi_{ij} - \frac{1}{3}\tr(\pi))\delta_{ij}  - \frac{1}{2} (A^{-1}x)_m (\epsilon_{mjk} \Omega_{ki} + \epsilon_{mik} \Omega_{kj})\right)\label{omegadot}
\end{eqnarray}
where the lapse function $N$ has been computed in (\ref{lapse expression}). These two equations completely define the dynamics.

\subsubsection{Second order formulation}

Finally, we would like to show explicitly the equality between the above Hamiltonian equations
and the second  order equations of motion 
written in \cite{ms1}.  To do so, we need to express $\Omega$ in terms of $\pi$ and $\dot{\pi}$ from the first equation (\ref{pidot}), and to implement this solution in the second equation
(\ref{omegadot}) in order to write a second order equation for $\pi$.

The expression of $\Omega$ as a function of $\pi$ and $\dot{\pi}$ is immediate. Indeed, multiplying (\ref{pidot}) on the left by $\pi^{-1}$ leads to 
\begin{eqnarray}
\Omega & = & -\frac{1}{2N}\left( \dot{\pi}\pi^{-1} + \pi^{-1}\dot{\pi}\right) \nonumber\\
\Gamma & = & -\frac{1}{2N} \left( \dot{\pi}\pi^{-1} - \pi^{-1}\dot{\pi} \right) \;\;\text{with} \;\; \Gamma_{ij}=\frac{1}{2}\epsilon_{ijm}(A^{-1}x)_m.\nonumber
\end{eqnarray}
On using these in (\ref{omegadot}) leads immediately to
the second order equation of motion
\begin{eqnarray}
\frac{1}{N} \dot{\Omega} + [\Gamma,\Omega] = m^2(\pi -\frac{1}{3}\tr(\pi))
\end{eqnarray}
which is exactly the same as the equation (66)  in \cite{ms1}. Furthermore, it is straightforward to show that the constraints $H\simeq 0$ and $\Psi \simeq 0$ are expressed in terms of $\pi$ and $\dot{\pi}$, they are equivalent
to the constraints (57) and (58) in \cite{ms1}, namely
\begin{eqnarray}
\Psi & = & \frac{m^2}{N\vert \pi \vert} \tr(\dot{\pi}\pi^{-1}) \simeq 0\nonumber \\
H & \simeq & -\frac{1}{4N^2} \tr((\dot{\pi} \pi^{-1} + \pi^{-1}\dot{\pi})^2) + 2m^2 (3 - \tr(\pi)) \simeq 0 .\nonumber
\end{eqnarray}
Thus, in this particular example, we have shown that, as expected, the first order dynamical system totally equivalent to the second order dynamical system studied in \cite{ms1}.

\section{Discussion}
\label{section7}

In this paper we have studied the canonical structure of the massive gravity in the first order moving frame formalism.  After recalling the main steps and the main results of the  Hamiltonian analysis of General relativity, we constructed the most general mass terms for massive gravity and carried out their ADM analysis. On working in the simplified context of translation invariant fields, we concluded that unless the mass terms take the specific dRGT form, namely linear in the lapse, the Boulaware-Deser ghost is present.  Thus we can also conclude that in the general case of space- and time-dependent fields, there will be additional degrees of freedom when the mass term is not linear in the lapse.  Then we carried out a complete Hamiltonian analysis of the dRGT massive gravity for translation invariant fields.  In this formalism, we have seen the origin of symmetry condition on $E$ (often assumed without proof, and required for the mass term to exist in the metric formulation.  Finally we determined the equations of motion through the calculation of the Dirac bracket, and thus obtained the time evolution of all the dynamical variables.  We checked that in certain specific cases,  in particular $\beta_0 \neq 0$, $\beta_1 \neq 0$, these reduces to those obtained from a totally different and much less general Lagrangian approach in \cite{ms1,ms2}.

We plan to continue this study in the future, and in particular focus in more detail on the time evolution and its well-posedness (which is not clear in the case of $\beta_3$ as seen in \cite{ms2}). This will be particularly important in the $\beta_2$ case, unstudied so far, but for which we now have the equations of motion. Finally, we can use the formalism developed here to determine new exact solutions of time-dependent translation-invariant massive gravity.

 \section*{Acknowledgments}
DAS is grateful to CERN for hospitality during the time the majority of this work was done.

\appendix

\section{Basis of Lorentz invariant functions of the tetrad fields}
\label{invariantfunctionsappendix}

In this appendix we express the basis functions 
\begin{eqnarray}
{\varphi}_0&=  &  \vert \theta \vert (N-N^a\theta_a^i\chi_i),\nonumber \\
{\varphi}_1& = &  N + \tr(\theta), \nonumber  \\ 
\varphi_2 &=&N \tr(\theta) - \theta^0_a \theta^a_0 + \frac{1}{2} \left[ (\tr(\theta))^2 - \tr(\theta^2)\right] 
 \nonumber 
\\
\varphi_3 &=&N \left[ (\tr(\theta))^2 - \tr(\theta^2) \right] + \frac{2}{3} \tr(\theta^3) - \tr(\theta)\tr(\theta^2)
+\frac{1}{3} (\tr(\theta))^3 + 2 \theta^0_a \theta^a_b \theta^b_0 - 2 \tr(\theta) \theta_a^0 \theta^a_0
\nonumber
\end{eqnarray}
defined in section \ref{section3}, in terms of the variables $E$, $\cal N$ and ${\cal N}^a$. To do so,  recall that
\begin{eqnarray}
\theta=\vert E \vert^{1/2} E^{-1} \;\;\;, \;\;\;
N= {\cal N} + {\cal N}^a \theta_a^i \chi_i  \;\;\;\text{and}\;\;\;
N^a={\cal N}^a + \frac{E^a_i \chi^i}{\vert E \vert^{1/2}} {\cal N},
\nonumber
\end{eqnarray}
and note the following useful relations
\begin{eqnarray}
\theta^0_a \theta^a_0 & = & \vert E \vert {\cal N}^j (E^{-2})^i_j \chi_i + \vert E \vert^{1/2} {\cal N} \chi^j (E^{-1})^i_j \chi_i \nonumber\\
\theta^0_a \theta^a_b \theta^b_0 & = & \vert E \vert^{3/2} {\cal N}^j (E^{-3})^i_j \chi_i + \vert E \vert {\cal N} \chi_i (E^{-2})^i_j \chi^j\nonumber \\
N-N^a \theta_a^i \chi_i & = & (1-\chi^2){\cal N}.
\nonumber
\end{eqnarray}
Then we find
\begin{eqnarray}
\varphi_n = \alpha_n {\cal N} +  {\cal N}^j  (M_n)^i_j \chi_i+ V_n \, ,  \qquad \forall \; n \in \{0,1,2,3 \}.
\nonumber
\end{eqnarray}
The functions $\alpha_n$ are given by
\begin{eqnarray}
&&\alpha_0 = \vert E \vert^{1/2} (1-\chi^2) \;\;\;,\;\;\; \alpha_1=1 \;\;\;,\;\;\; \alpha_2=\vert E \vert^{1/2} \left( \tr(E^{-1}) - \chi^j (E^{-1})^i_j \chi_i \right)    \nonumber\\
&&\alpha_3=\vert E \vert \left(  (\tr(E^{-1}))^2 - \tr(E^{-2}) + 2 \chi^j (E^{-2})^i_j \chi_i - 2 \tr(E^{-1}) \chi^j(E^{-1})^i_j \chi_i  \right).
\nonumber
\end{eqnarray}
The functions $V_n$ are given by
\begin{eqnarray}
&& V_0 =0  \;\;\;,\;\;\; V_1=\vert E \vert^{1/2} \tr(E^{-1})  \;\;\;,\;\;\; 
V_2=\frac{1}{2}\vert E \vert \left( (\tr(E^{-1}))^2 - \tr(E^{-2}) \right)  \nonumber \\
&&V_3=\vert E \vert^{3/2} \left( \frac{2}{3} \tr(E^{-3}) - \tr(E^{-1}) \tr(E^{-2}) + \frac{1}{3} (\tr(E^{-1}))^3 \right).
\nonumber
\end{eqnarray}
Finally, the matrices $M_n$ are defined by
\begin{eqnarray}
&&M_0=0  \;\;\;,\;\;\;  M_1=\vert E \vert^{1/2}E^{-1} \;\;\;,\;\;\;
M_2=\vert E \vert \left( (\tr(E^{-1}))E^{-1} - E^{-2}\right) \nonumber \\
&&M_3=\vert E \vert^{3/2} \left( ((\tr(E^{-1}))^2 - \tr(E^{-2})) E^{-1} - 2 \tr(E^{-1}) E^{-2} + 2 E^{-3}  \right).
\nonumber
\end{eqnarray}

\section{Hamiltonian form of the dRGT mass terms}
\label{appendix2}

In this section, we detail the steps required to obtain the Hamiltonian form of the dRGT mass terms (\ref{ll0})-(\ref{ll3}).

The cosmological constant term (\ref{ll0}) is given by
\begin{eqnarray}
{\cal L}_0 & = & \frac{\beta_0}{4!} \epsilon_{IJKL} \epsilon^{\mu\nu\rho\sigma} \theta_\mu^I \theta_\nu^J \theta_\rho^K \theta_\sigma^L =
                  \frac{\beta_0}{3!} \epsilon_{ijk} \epsilon^{abc} \left( \theta_0^0 \theta_a^i  \theta_b^j  \theta_c^k - 3  \theta_a^i \theta_b^j  \theta_0^k  \theta_c^0 \right) \nonumber \\
                  &=& \beta_0 \vert E \vert^{1/2} \left( N - N^a \theta_a \cdot \chi\right) = \beta_0 \vert E \vert^{1/2} (1-\chi^2) {\cal N},
\end{eqnarray}
while (\ref{ll1}) takes the form
\begin{eqnarray}
{\cal L}_1 & = & \frac{\beta_1}{3!}  \epsilon_{IJKL} \epsilon^{\mu\nu\rho\sigma} f_\mu^I \theta_\nu^J \theta_\rho^K \theta_\sigma^L
 =  \frac{\beta_1}{ 3!}  \epsilon_{ijk} \epsilon^{abc} \left( 3N f_a^i \theta_b^j \theta_c^k - 6N^d f_a^i \theta_b^j \theta_c^\ell \theta_d^k \chi_\ell  + \theta_a^i \theta_b^j \theta_c^k \right) \nonumber \\
                 & = &  \beta_1 \left( N \text{tr}(fE)  + \vert E \vert^{1/2}   - N^b[(\theta_b\cdot \chi)\text{tr}(fE) - (f_b \cdot \chi)\vert E \vert^{1/2}] \right)\nonumber \\
                 & = & \beta_1 \left( \vert E \vert^{1/2} + {\cal N}[(1-\chi^2)\text{tr}(fE) +(E^a\cdot \chi)(f_a \cdot \chi)  ] + \vert E \vert^{1/2}{\cal N}^a  (f_a \cdot \chi) \right) \nonumber \\
                & = & \beta_1 \left( \vert E \vert^{1/2} + {\cal N}[(1-\chi^2)\text{tr}(E) +\chi \cdot (E \chi)] + \vert E \vert^{1/2}{\cal N}^a\chi_a  \right)  ,
\end{eqnarray}
where we have used the notation $\tr(fE)=f_a^i E^a_i$ for the trace. Notice that, as $f$ enables us to identify space-time indices with internal indices,  we can extend the scalar product
notation $u \cdot v$ to denote the contraction between a one-form and a vector field.    Only in the last line have we set $f^I_\mu=\delta_\mu^I$ consistently with (\ref{fdelta}).

We proceed in the same way to find the expression of ${\cal L}_2$:
\begin{eqnarray}
{\cal L}_2 & = & \frac{\beta_2}{2} \epsilon_{IJKL} \epsilon^{\mu\nu\rho\sigma} f_\mu^I f_\nu^J \theta_\rho^K \theta_\sigma^L
 =  2{\beta_2} \epsilon_{ijk} \epsilon^{abc} \left( N f_a^i f_b^j \theta_c^k - N^d  f_a^i f_b^j \theta_d^k   \theta_c^\ell \chi_\ell + f_a^i   \theta_b^j   \theta_c^k \right) \nonumber \\
 & = &  2\beta_2 \left( N \tr(\theta) + \tr(E) - N^a(\theta^2)^i_a \chi_i \right).
 \label{theta2a}
 \nonumber
\end{eqnarray}
To go further, we use  the relation (\ref{inverse E}) between $\theta$ and $E^{-1}$ and also the Cayley Hamilton theorem for the (3 dimensional) matrix $\theta$ which states that
\begin{eqnarray}
\theta^2 = \tr(\theta) \theta + \vert \theta \vert \theta^{-1} - \vert \theta \vert \tr(\theta^{-1}) \mathbb I \;.
\label{Cayley}
\end{eqnarray}
This allows us to replace the $\theta^2$ in (\ref{theta2a}) and to obtain the following simplified expression of ${\cal L}_2$ :
\begin{eqnarray}
{\cal L}_2 = {2} \beta_2 \left( \tr(E) + {\cal N} \vert E \vert^{1/2} [\tr(E^{-1}) -\chi \cdot (E^{-1}\chi)] +{\cal N}^a[ (E\chi)_a - (\tr(E)) \chi_a]    \right).
\end{eqnarray}

We finish with the computation of ${\cal L}_3$ using the same strategy as in the previous cases:
\begin{eqnarray}
{\cal L}_3 & = &  \beta_3  \epsilon_{IJKL} \epsilon^{\mu\nu\rho\sigma} f_\mu^I f_\nu^J f_\rho^K \theta_\sigma^L
 =  \beta_3 \epsilon_{ijk} \epsilon^{abc} \left( N f_a^i f_b^j f_c^k + 3  f_a^i f_b^j \theta_c^k \right) \nonumber \\
                 & = & 6\beta_3(N + \tr(\theta)) = 6 \beta_3 \left( \vert E \vert^{1/2} \tr(E^{-1}) + {\cal N} + \vert E \vert^{1/2} {\cal N}^a (E^{-1}\chi)_a\right).
\end{eqnarray}

As a consequence, we end up with
\begin{eqnarray}
{\cal L}_m  = V +  \frac{\cal N}{2\vert E\vert^{1/2}} H^m + {\cal N}^a H_a^{m}
\end{eqnarray}
where the explicit form of $V$, $H^m$ and $H_a^m$ are given in section \ref{section5}.

\section{Poisson bracket between $\pi$ and $\Omega$ matrix elements}
\label{Poisson OmegaPi}
In this section, we compute the Poisson brackets between the variables $\pi=\vert E \vert^{-1/2}E$ and $\Omega=\pi A$ from the canonical Poisson bracket
\begin{eqnarray}
\{ E_{ij},A_{k\ell}\} = \delta_{jk} \, \delta_{i\ell} .
\nonumber
\end{eqnarray}
We will show that
\begin{eqnarray}
\{ \pi_{ij},\pi_{k\ell}\}& = & 0 \label{PoissonPi}\\
\{ \pi_{ij},\Omega_{k\ell}\} & = & \vert \pi \vert (  \delta_{i \ell} \pi_{kj} - \frac{1}{2} \delta_{k\ell} \pi_{ij} ) \label{remainPoisson1} \\
\{ \Omega_{ij},\Omega_{k\ell}\} & = & \vert \pi \vert (\delta_{i\ell} \Omega_{kj} - \delta_{jk} \Omega_{i\ell} -\frac{1}{2}\delta_{k\ell} \Omega_{ij} + \frac{1}{2}\delta_{ij} \Omega_{k\ell}) \label{remainPoisson2}. 
\end{eqnarray}
The first identity (\ref{PoissonPi}) is straightforward 
because $\pi$ depends only on $E$. To compute the two remaining Poisson brackets we first need to establish useful formulae. We start with the following one:
\begin{eqnarray}
\{ A_{ij}, \vert E \vert \} & = & \{ A_{ij} , \frac{1}{6} \epsilon^{abc} \epsilon^{k\ell m} E_{ka} E_{\ell b} E_{mc} \}= \frac{1}{2}   \epsilon^{abc} \epsilon^{k\ell m} E_{ka} E_{\ell b} \{ A_{ij}, E_{mc}\} \nonumber\\
 & = & -{\vert E \vert} E^{-1}_{mc} \, \delta_{j}^m \, \delta_{i}^c = - \vert E \vert E^{-1}_{ij}\nonumber
\end{eqnarray}
where we used the definition of the determinant for the $3\times 3$ (and invertible)  matrix $E$
\begin{eqnarray}
\vert E \vert = \frac{1}{6}  \epsilon^{abc} \epsilon^{ijk} E_{ia} E_{j b} E_{kc} \;\;\; \Longrightarrow  \;\;\; E^{-1}_{ia} = \frac{1}{2\vert E \vert} \epsilon^{ijk} \epsilon^{abc} E_{jb} E_{kc}.
\nonumber
 \end{eqnarray}
From these formulae, we can simplify the expression of the following Poisson bracket
\begin{eqnarray}\label{formula1}
\{ \pi_{ij},A_{k\ell}\}  =  \vert E \vert^{-1/2} \delta_{jk} \delta_{i\ell} + E_{ij} \{ \vert E \vert^{-1/2}, A_{k\ell}\} 
  =  \vert E \vert^{-1/2} \delta_{jk} \delta_{i\ell} -\frac{ E_{ij}  E^{-1}_{k\ell}}{2 \vert E \vert^{1/2}}  .
\end{eqnarray}
Now, we have all the ingredients to compute (\ref{remainPoisson1}) and (\ref{remainPoisson2}). Then (\ref{remainPoisson1} follows immediately
\begin{eqnarray}\label{PoissonPiOmega}
\{ \pi_{ij},\Omega_{k\ell}\} = \pi_{kb} \{ \pi_{ij},A_{b\ell}\} = \vert \pi \vert (  \delta_{i \ell} \pi_{kj} - \frac{1}{2} \delta_{k\ell} \pi_{ij} ).
\end{eqnarray}
The second Poisson bracket (\ref{remainPoisson2}) is obtained as follows
\begin{eqnarray}
\{ \Omega_{ij},\Omega_{k\ell}\} & = & \{ \pi_{ia}A_{aj},\pi_{kb}A_{b\ell}\} =  A_{aj} \{ \pi_{ia},\pi_{kb}A_{b\ell}\} \pi_{kb} + \pi_{ia} \{A_{aj},\pi_{kb}\} A_{b\ell} \nonumber\\
& = & A_{aj} \vert E \vert^{-1/2}  (\delta_{i\ell} \delta_{ab} - \frac{1}{2}E_{ia}E^{-1}_{b\ell}) \pi_{kb}+ \pi_{ia} \vert E \vert^{-1/2}(-\delta_{kj}\delta_{ab} + \frac{1}{2} E_{kb}E^{-1}_{aj}) A_{b\ell} \nonumber\\
& = & \vert \pi \vert (\delta_{i\ell} \Omega_{kj} - \delta_{jk} \Omega_{i\ell} -\frac{1}{2}\delta_{k\ell} \Omega_{ij} + \frac{1}{2}\delta_{ij} \Omega_{k\ell})
\nonumber
\end{eqnarray}
where we used (\ref{formula1}) in the second line, the definitions  $\Omega=\pi A$ and $\pi=\vert E \vert^{-1/2}E$ and the relation between the determinants $\vert E \vert^{-1/2}=\vert \pi \vert$ in the third line.

\section{Explicit calculation of the Dirac matrix}
\label{appendix Dirac}
This appendix is devoted to details of the calculation of the Dirac matrix, i.e.~the Poisson bracket between the second class constraints.  
To do so, we first need to establish the following formulae which hold at least weakly:
\begin{eqnarray}
\{ \pi^{-1}_{ij},\Omega_{k\ell}\} & \simeq & -\vert \pi \vert (\delta_{jk} \pi^{-1}_{i\ell} - \frac{1}{2}\delta_{k\ell} \pi^{-1}_{ij}) \label{inversePi}\\
\{ \vert \pi \vert, \Omega_{ij} \} & \simeq &  -\frac{1}{2}\vert \pi \vert^2 \delta_{ij} \label{detOmega}\\
\{\pi_{ij},\tr(\Omega) \} & \simeq & -\frac{1}{2} \vert \pi \vert \pi_{ij} \label{traceO}\\
\{ \pi_{ij},\tr(\Omega^2)\} & \simeq & 2\vert \pi \vert ( (\Omega \pi)_{ij} - \frac{1}{2} \tr(\Omega) \pi_{ij}) \label{traceOO}\\
\{ \pi_{ij},\tr(\Omega\pi)\} & \simeq &\vert \pi \vert (\pi^2_{ij} - \frac{1}{2} \tr(\pi) \pi_{ij})  \label{tracePiOPi}\\
\{ \pi_{ij},\tr(\Omega\pi^{-1})\} & \simeq &\vert \pi \vert (\delta_{ij} - \frac{1}{2} \tr(\pi^{-1}) \pi_{ij}) \label{tracePiOPi-1} \\
\{ \Omega_{ij},\tr(\Omega)\} & \simeq & \frac{\vert \pi \vert}{2} (\tr(\Omega)\delta_{ij} - 3 \Omega_{ij}) \label{O1}\\
\{ \Omega_{ij},\tr(\Omega^2)\} & \simeq & \vert \pi \vert (\tr(\Omega^2)\delta_{ij} - \tr(\Omega) \Omega_{ij}) \label{O2}\\
\{ \Omega_{ij},\tr(\pi)\} & \simeq & - \vert \pi \vert (\pi_{ij} - \frac{1}{2} \tr(\pi) \delta_{ij}) \label{O3}\\
\{ \Omega_{ij}, \tr(\pi^{-1})\} & \simeq & \vert \pi \vert (\pi^{-1}_{ij} - \frac{1}{2} \tr(\pi^{-1}) \delta_{ij}). \label{O4}\\
\{ \Omega_{ij},\tr(\Omega\pi)\} & \simeq &\vert \pi \vert (\tr(\Omega\pi) \delta_{ij} -(\Omega\pi)_{ij} - \frac{1}{2} \tr(\pi) \Omega_{ij}) \label{traceOPi}\\
\{ \Omega_{ij},\tr(\Omega\pi^{-1})\} & \simeq &\vert \pi \vert ((\pi^{-1}\Omega)_{ij} - \frac{1}{2} \tr(\pi^{-1}) \Omega_{ij}) \label{traceOPi-1}
\end{eqnarray}
They follow immediately from the Poisson brackets computed in the previous appendix.

\medskip

To compute the Dirac matrix and also the Dirac bracket, we need to know the Poisson brackets between the matrix elements of $\pi$ and $\Omega$ with the constraints.
Using all the previous formulae, it is long but straightforward to obtain the following results which hold at least weakly:
\begin{eqnarray}
\{ C_m,\pi_{ij}\} & \simeq & 0\nonumber \\
\{ D_m,\pi_{ij} \} & \simeq & \vert \pi \vert \epsilon_{mik} \pi_{kj} \nonumber\\
\{H,\pi_{ij} \} & \simeq & 2\vert \pi \vert (\Omega\pi)_{ij} \nonumber\\
\{\Psi,\pi_{ij} \} & \simeq & -6\beta_3 \vert \pi \vert \delta_{ij} + (\frac{1}{2}\beta_1 - \beta_2 \tr(\pi) + 3\beta_3 \vert \pi \vert \tr(\pi^{-1})) \pi_{ij} + 2\beta_2 \pi^2_{ij} \nonumber\\
\{C_m,\Omega_{ij} \} & \simeq & \vert \pi \vert \epsilon_{mjk}\pi_{ki} \nonumber\\
\{D_m,\Omega_{ij} \} & \simeq & \vert \pi \vert (\epsilon_{mjk}\Omega_{ki} + \epsilon_{mik}\Omega_{kj}) \nonumber\\
\{H,\Omega_{ij} \} & \simeq & \vert \pi \vert (\tr(\Omega^2) - (\tr\Omega)^2 - \beta_1 \tr(\pi) -6\beta_3 \vert \pi \vert ) \delta_{ij} \nonumber \\
 &&+2\vert \pi \vert \tr(\Omega) \Omega_{ij} + 2\vert \pi \vert \beta_1 \pi_{ij} - 8\vert \pi \vert^2 \beta_2 \pi^{-1}_{ij} \nonumber\\
 \{\Psi,\Omega_{ij} \} & \simeq & \beta_2\tr(\Omega\pi) \delta_{ij} +(\frac{3}{2}\beta_1 - \beta_2 \tr(\pi) + 3\beta_3 \vert \pi \vert \tr(\pi^{-1})) \Omega_{ij} \nonumber \\
 && -2\beta_2 (\Omega\pi)_{ij} -6\beta_3\vert\pi\vert (\pi^{-1}\Omega)_{ij}\nonumber
\end{eqnarray}

From these results, computing the Poisson brackets between the different constraints becomes immediate. Some of them are in fact vanishing at least weakly:
\begin{eqnarray}
\{ C_m,C_n\} = 0 \;,\;\;\;  \{C_m,\Psi\}\simeq 0 \;,\;\; \{D_m,H\}\simeq 0 \;,
\nonumber
\end{eqnarray} 
and the only non trivial Poisson brackets are the remaining ones
 \begin{eqnarray}
 \{C_m,D_n\} & \simeq & A_{mn} \;\;\; \text{with} \;\;\; A=\vert \pi \vert (\pi - \tr(\pi))  \nonumber\\
 \{H,C_m\} & \simeq & x_m \;\;\; \text{with} \;\;\; x_m= 2\vert \pi \vert \epsilon_{mij} (\Omega\pi)_{ij}\nonumber \\
 \{\Psi,D_m\} & \simeq & y_m \;\;\; \text{with} \;\;\; y_m= -2 \epsilon_{mij}(\beta_2 \Omega\pi + 3\beta_3 \vert \pi \vert \pi^{-1}\Omega)_{ij}\nonumber\\
 \{\Psi,H\} & \simeq & a \nonumber
 \end{eqnarray}
 where $a$ is a scalar function in the phase space whose explicit form is computed below here. This gives the form of the Dirac matrix given in the core of the article.
 
 To finish, let us compute the expression of $a$. For that purpose, we need the following Poisson brackets between scalar functions in the phase space:
 \begin{eqnarray}
 \{\vert \pi \vert, \tr(\Omega)\} & \simeq & -\frac{3}{2} \vert \pi \vert^2 \nonumber\\
 \{\vert \pi \vert , \tr(\Omega^2)\} & \simeq & -\vert \pi \vert^2 \tr(\Omega) \nonumber\\
 \{\tr(\Omega) ,\tr(\Omega^2) \} & \simeq & \vert \pi \vert (3\tr(\Omega^2) - (\tr\Omega)^2)\nonumber\\
 \{\tr(\pi) , \tr(\Omega)\} & \simeq & -\frac{1}{2} \vert \pi \vert \tr(\pi) \nonumber\\
  \{\tr(\pi) ,\tr(\Omega^2) \} & \simeq & \vert \pi \vert (2\tr(\Omega\pi) -\tr(\Omega)\tr(\pi))\nonumber \\
  \{ \tr(\pi^{-1}),\tr(\Omega) \} & \simeq & \frac{1}{2} \vert \pi \vert \tr(\pi^{-1}) \nonumber\\
  \{ \tr(\Omega\pi), \tr(\Omega) \} & \simeq & \vert \pi \vert (\frac{1}{2} \tr(\pi)\tr(\Omega) - 2\tr(\Omega\pi)) \nonumber\\
  \{ \tr(\Omega\pi), \tr(\Omega^2) \} & \simeq & \vert \pi \vert (2\tr(\Omega^2\pi) + \tr(\pi) \tr(\Omega^2) - 2 \tr(\Omega) \tr(\Omega\pi))\nonumber \\
  \{ \tr(\Omega\pi), \tr(\pi)\} & \simeq & \vert \pi \vert (\frac{1}{2} (\tr\pi)^2 -\tr(\pi^2)) \nonumber\\
   \{\tr(\Omega\pi) ,\vert \pi \vert \} & \simeq & \frac{1}{2} \vert \pi \vert^2 \tr(\pi)\nonumber \\
   \{ \tr(\Omega\pi) , \tr(\pi^{-1}) \} & \simeq & \vert \pi \vert (3 -\frac{1}{2} \tr(\pi) \tr(\pi^{-1}))\nonumber\\
    \{ \tr(\Omega\pi^{-1}),\tr(\Omega) \} & \simeq & \vert \pi \vert (\frac{1}{2} \tr(\pi^{-1}) \tr(\Omega) -\tr(\pi^{-1} \Omega)) \nonumber\\
   \{  \tr(\Omega\pi^{-1}),\tr(\Omega^2) \} & \simeq & \vert \pi \vert (\tr(\pi^{-1}) \tr(\Omega^2) -2\tr(\pi^{-1} \Omega^2))\nonumber \\
   \{ \tr(\Omega\pi^{-1}),\tr(\pi) \} & \simeq & \vert \pi \vert (\frac{1}{2} \tr(\pi)\tr(\pi^{-1}) -3) \nonumber\\
   \{ \tr(\Omega\pi^{-1}), \vert \pi \vert \} & \simeq & \frac{1}{2} \vert \pi \vert^2 \tr(\pi^{-1}) \nonumber\\
   \{ \tr(\Omega\pi^{-1}), \tr(\pi^{-1}) \} & \simeq & \vert \pi \vert (\tr(\pi^{-2}) - \frac{1}{2} \tr(\pi^{-1})^2)            \nonumber
 \end{eqnarray}
All these results enter in the calculation of $a$ and after a long but immediate calculation we obtain the following expression:
\begin{eqnarray}
a & \simeq & 3\beta_1 \left((\tr\Omega)^2 - \tr(\Omega^2)\right) 
- 4 \beta_2 \left(  \tr(\Omega^2\pi) + \tr(\pi) \tr(\Omega^2) + 2\tr(\Omega) \tr(\Omega\pi)   \right) \nonumber \\
  & + & 6\beta_3 \vert \pi \vert \left( \tr(\pi^{-1}) (\tr\Omega)^2 - 2 \tr(\Omega) \tr(\pi^{-1} \Omega) + 2\tr(\pi^{-1} \Omega^2) -\tr(\pi^{-1}) \tr(\Omega^2)\right) \nonumber \\
  & + & \beta_1^2 \tr(\pi) -48 \beta_2^2 \vert \pi \vert + 36 \beta_3^2 \vert \pi \vert^2 \tr(\pi^{-1}) + 2 \beta_1 \beta_2 \left(4\vert \pi \vert \tr(\pi^{-1}) -(\tr\pi)^2 + 2\tr(\pi^2)\right) \nonumber \\
  & + & 6\beta_1\beta_3 \vert \pi \vert \left(\tr(\pi) \tr(\pi^{-1}) - 3\right) +12\beta_2\beta_3 \vert \pi \vert \left(4\vert \pi \vert \tr(\pi^{-2})- \tr(\pi)\right).
  \nonumber
\end{eqnarray}

\end{document}